\documentclass[sigconf,10pt,screen]{acmart}

\definecolor{brightpink}{rgb}{1.0, 0.0, 0.5}
\definecolor{awesome}{rgb}{1.0, 0.13, 0.32}

\usepackage[normalem]{ulem}

\usepackage{graphicx}
\usepackage{listings}
\usepackage{epstopdf}
\usepackage{color}
\usepackage{subfigure}
\usepackage{xspace}
\usepackage{latexsym}
\makeatletter
\newif\if@restonecol
\makeatother

\usepackage[ruled,vlined,linesnumbered]{algorithm2e}
\SetAlFnt{\small}
\usepackage{algorithmic}

\SetCommentSty{mycommfont}
\SetKwComment{Comment}{$\triangleleft$\ }{}

\usepackage{enumerate}

\usepackage{wrapfig}
\usepackage{setspace}
\usepackage{blindtext}
\usepackage{amsfonts}
\usepackage{multirow}
\usepackage{pifont}
\usepackage{mathrsfs}
\usepackage{mathtools}
\usepackage{relsize}
\usepackage{ulem}
\usepackage{comment}
\usepackage{footnote}
\makesavenoteenv{tabular}
\makesavenoteenv{table}
\usepackage[frozencache,cachedir=.]{minted}

\usepackage[font=small]{caption}

\usepackage{booktabs}
\usepackage{siunitx}

\usepackage{microtype}

\usepackage[hyphenbreaks]{breakurl} 
\usepackage{minted}
\usepackage{cleveref}
\crefformat{section}{\S#2#1#3}

\newif{\ifSubmit}
\newif{\ifFinal}
\newif{\ifDraft}
\Drafttrue

\ifDraft
\newlength{\gapspace}

\newcommand{\bencomment}[1]{\noindent\textcolor{blue}{\bf Ben: #1}}

\newcommand{\aocomment}[1]{\noindent\textcolor{green}{\bf Ao: #1}}
\newcommand{\yuecomment}[1]{\noindent\textcolor{red}{\bf Yue: #1}}

\fi

\newcommand{\proj}{\textsc{Wukong}}

\setcopyright{rightsretained}

\copyrightyear{2020} 
\acmYear{2020} 
\setcopyright{acmcopyright}\acmConference[SoCC '20]{ACM Symposium on Cloud Computing}{October 19--21, 2020}{Virtual Event, USA}
\acmBooktitle{ACM Symposium on Cloud Computing (SoCC '20), October 19--21, 2020, Virtual Event, USA}
\acmPrice{15.00}
\acmDOI{10.1145/3419111.3421286}
\acmISBN{978-1-4503-8137-6/20/10}

\begin{document}

\title{
{\proj}: A Scalable and Locality-Enhanced Framework for Serverless Parallel Computing
}

\author{Benjamin Carver}
\affiliation{%
  \institution{George Mason University}
}
\email{bcarver2@gmu.edu}

\author{Jingyuan Zhang}
\affiliation{%
  \institution{George Mason University}
}
\email{jzhang33@gmu.edu}

\author{Ao Wang}
\affiliation{%
  \institution{George Mason University}
}
\email{awang24@gmu.edu}

\author{Ali Anwar}
\affiliation{%
  \institution{IBM Research--Almaden}
}
\email{Ali.Anwar2@ibm.com}

\author{Panruo Wu}
\affiliation{%
  \institution{University of Houston}
}
\email{pwu7@uh.edu}

\author{Yue Cheng}
\affiliation{%
  \institution{George Mason University}
}
\email{yuecheng@gmu.edu}

\begin{abstract}
Executing complex, burst-parallel, directed acyclic graph (DAG) jobs poses a major challenge for serverless execution frameworks, which will need to rapidly scale and schedule tasks at high throughput, while minimizing data movement across tasks. We demonstrate that, for serverless parallel computations, decentralized scheduling enables scheduling to be distributed across Lambda executors that can schedule tasks in parallel, and brings multiple benefits, including enhanced data locality, reduced network I/Os, automatic resource elasticity, and improved cost effectiveness.
We describe the implementation and deployment of our new serverless parallel framework, called {\proj},
on AWS Lambda. We show that {\proj} achieves near-ideal scalability, executes parallel computation jobs up to $68.17\times$ faster, reduces network I/O by multiple orders of magnitude, and achieves $92.96\%$ tenant-side cost savings compared to numpywren.
\end{abstract}

\maketitle

\vspace{-5pt}
\section{Introduction}
\label{sec:intro}

In recent years, a new cloud computing model called serverless
computing or Function as a Service (FaaS)~\cite{aws_serverless}
has emerged. Serverless computing enables a new way of building and scaling applications and services by allowing developers to break traditionally monolithic server-based
applications into finer-grained cloud functions.
Developers write function logic while the service provider performs the notoriously tedious tasks of provisioning, scaling, and managing the backend servers that the functions run on \cite{gray_stop}.

Serverless computing solutions are growing in popularity and finding their way into both commercial clouds (e.g., AWS Lambda, Google Cloud Functions, and Azure Functions)
and open source projects (e.g., OpenWhisk). 
While serverless platforms were originally intended for event-driven, stateless applications~\cite{serverless_usecases},
a recent trend is the use of serverless computing for more complex, stateful, parallel applications.  

Some types of compute- and data-intensive applications are inherently parallelizable and can be
structured as a directed acyclic graph (DAG) of short, fine-grained tasks~\cite{sparrow_sosp13, alibaba_trace18, alitrace_iwqos19, gg_atc19}.
The large-scale parallelism and auto-scaling services provided by serverless platforms makes them well-suited for such kinds of burst-parallel fine-grained tasks 
that characterize DAG-based parallel computation workflows. 
Burst-parallel applications include data analytics~\cite{alitrace_iwqos19}, optimization algorithms~\cite{convex_dask}, and real-time machine learning classifications such as support vector machines (SVM)~\cite{svn, scikit_svm, psvm_nips07}; these applications typically demand low-latency scheduling~\cite{sparrow_sosp13} with large-scale parallelism~\cite{berkeley_parallel_computing}.

\begin{figure}[t]
\begin{center}
\includegraphics[width=0.4\textwidth]{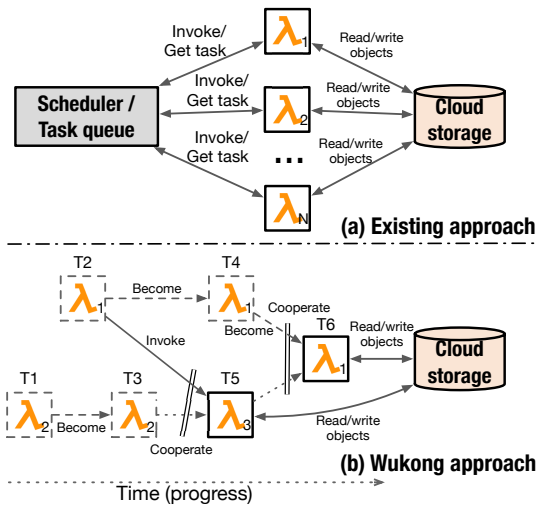}
\vspace{-10pt}
\caption{
In (a), the central scheduler tracks all task completions, updates all task dependencies, and identifies all ready tasks. The scheduler dispatches ready tasks to Lambda executors; or the scheduler deposits ready tasks into a shared work queue, and a pool of Lambda executors contend for the queued tasks. Intermediate task inputs and outputs are stored outside of the executors, which reduces data locality. In (b), task scheduling is performed by a fleet of Lambda executors that schedule and execute their assigned tasks in parallel and cooperate to ensure that task dependencies are satisfied. Intermediate task inputs and outputs may be stored inside the executors, which increases data locality. 
This approach also enables fine-grained and automatic Lambda resource elasticity, as Lambda executors finish assigned tasks and return (e.g., Lambda 2). 
}
\vspace{-20pt}
\label{fig:design_diff}
\end{center}
\end{figure}

FaaS providers charge function execution time at a fine granularity -- 
AWS Lambda bills on a per-invocation basis.

Workloads with short tasks can take advantage of this fine-grained pay-per-use pricing model to keep monetary costs low.
Consequently, serverless computing can be leveraged by next-generation, burst-parallel workloads in high-performance computing (HPC) and data analytics.

Migrating such applications from a traditional serverful deployment to a serverless platform
presents unique opportunities. Traditional serverful deployments rely on existing workflow management frameworks such as MapReduce~\cite{mapreduce_osdi04}, Apache Spark~\cite{spark_nsdi12},  Sparrow~\cite{sparrow_sosp13}, and Dask~\cite{dask} to provide a logically centralized scheduler for managing task assignments and resource allocation. The scheduler traditionally has various objectives, including load balancing, maximizing cluster utilization, ensuring task fairness, and so on. 
However, a traditional serverful scheduler is not required by
serverless computing.
This is because: (1) FaaS providers are responsible for managing the ``servers'' (i.e., where the task executors are hosted); and (2) serverless platforms typically provide a \emph{nearly unbounded} amount of \emph{ephemeral} resources. As a result, a hypothetical serverless parallel computing framework may not necessarily care about traditional ``scheduling''-related metrics (such as load balancing and cluster utilization), since the framework has no control over where tasks are executed. (The service provider, of course, cares about these metrics.)

Yet, designing an efficient serverless-oriented parallel computing framework introduces unique challenges. 
First, while serverless platforms (e.g., AWS Lambda) promise to offer superior elasticity and auto-scaling properties, the serverless invocation model imposes
non-trivial scheduling overhead. Unlike a typical serverful parallel framework where the central scheduler directly communicates with each worker process using TCP~\cite{mapreduce_osdi04, hadoop}, in a serverless setup, the scheduler can dispatch tasks to serverless workers in one of three ways.

Figure~\ref{fig:design_diff}(a) depicts a high-level overview of all three methods.
In method \#1, the scheduler invokes a Lambda function (using the HTTP protocol) to dispatch the task code and execute the task. Note that with this method, there is a one-to-one association between tasks and Lambda functions. 
Given an average invocation overhead of 50 ms (typical for AWS Lambda functions), the scheduler could quickly become a performance bottleneck, especially for large and complex jobs with thousands of tasks. These observations indicate that a naive attempt to simply port an existing serverful DAG framework to serverless computing will be unsuccessful. In order to create a performant, cost-effective serverless DAG engine, new techniques must be developed to fully take advantage of the characteristics of the serverless platform.

In method \#2, the scheduler launches short-lived\footnote{Lambda functions may run up to 900 seconds in AWS cloud.} Lambda executors as workers that establish TCP connections with the scheduler and receive RPC requests for task processing. Example frameworks include ExCamera~\cite{excamera_nsdi17} and PyWren~\cite{pywren_socc17}. Task executors within these frameworks may execute several tasks as opposed to just one as with the first method.

Similarly, in method \#3, the scheduler places tasks in a shared work queue. These tasks are retrieved from the queue by  serverless executors; state-of-the-art systems such as numpywren~\cite{numpywren} launch stateless Lambda executors that connect to a centralized shared queue and constantly retrieve tasks from the queue. 
Frameworks using this method may have a component separate from the central scheduler that is responsible for invoking the AWS Lambda executors. This is sometimes referred to as a ``provisioner''.
In the latter two approaches, as shown in Figure~\ref{fig:design_diff}(a),
a tightly synchronized central scheduler tracks all task completions, updates all task dependencies, and identifies any ready tasks. The scheduler dispatches ready tasks to Lambda executors, or the scheduler deposits ready tasks into a shared work queue, and a pool of Lambda executors contend for the queued tasks. 
Intermediate task inputs and outputs are stored externally, which reduces data locality.

The second challenge  is that serverless platforms come with inherent constraints, including bandwidth-limited, outbound only network connectivity; therefore, serverless workflows must rely on  external cloud store for intermediate data storage and exchange, which creates excessive data movement overhead. Data locality enhancement is thus critical for minimizing  communication costs.

Researchers have developed serverless parallel computing frameworks that support parallel job processing~\cite{pywren_socc17, excamera_nsdi17, numpywren}; however, these solutions do not fully address the aforementioned performance issues of efficient scheduling and data locality, which leads to long scale-out delays, sub-optimal performance, and higher monetary cost. To this end, we design and build a new serverless parallel computing framework called {\proj}\footnote{\url{https://mason-leap-lab.github.io/Wukong}}. {\proj} is a serverless-oriented, decentralized, locality-aware, and cost-effective parallel computing framework. \emph{The key insight of {\proj} is that partitioning the work of a centralized scheduler (i.e., tracking task completions, identifying and dispatching ready tasks, etc.) across a large number of Lambda executors, can greatly improve performance by permitting tasks to be scheduled in parallel, reducing resource contention during scheduling, and making task scheduling data locality-aware, with automatic resource elasticity and improved cost effectiveness.}

\begin{figure*}
\vspace{-15pt}
\begin{minipage}{\textwidth}
\begin{minipage}[b]{0.3\textwidth}
\begin{center}
\includegraphics[width=1\textwidth]{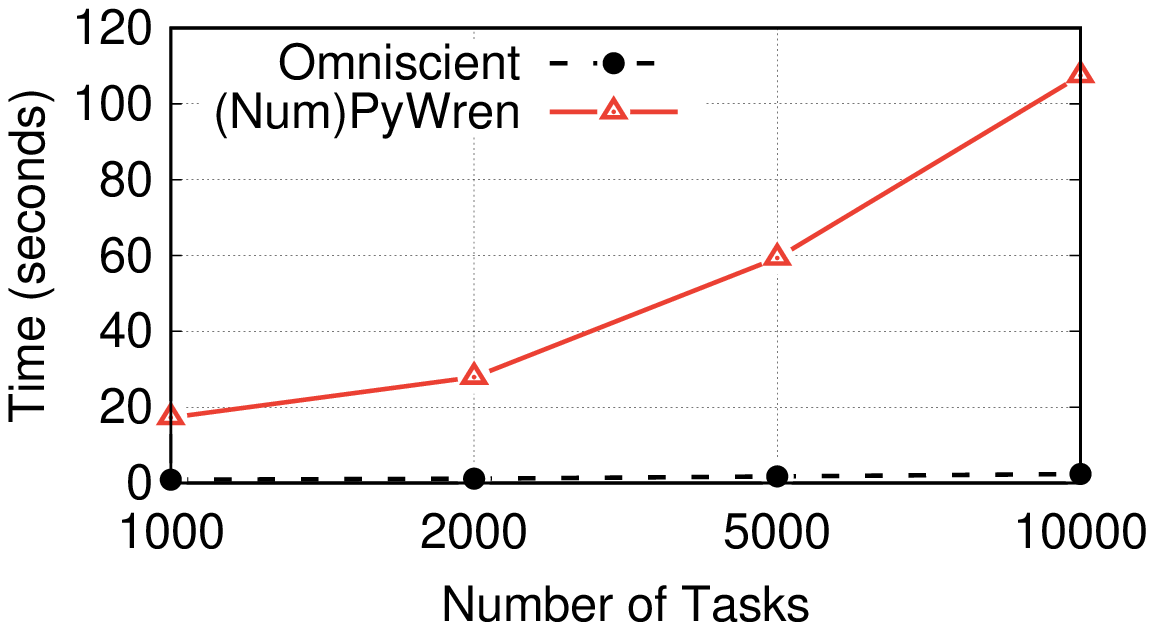}
\end{center}
\vspace{-25pt}
\caption{
\small{(Num)PyWren scaling tasks on AWS Lambda.}
}
\vspace{-8pt}
\label{fig:moti_pywren}
\end{minipage}
\hfill
\begin{minipage}[b]{0.3\textwidth}
\begin{center}
\includegraphics[width=1\textwidth]{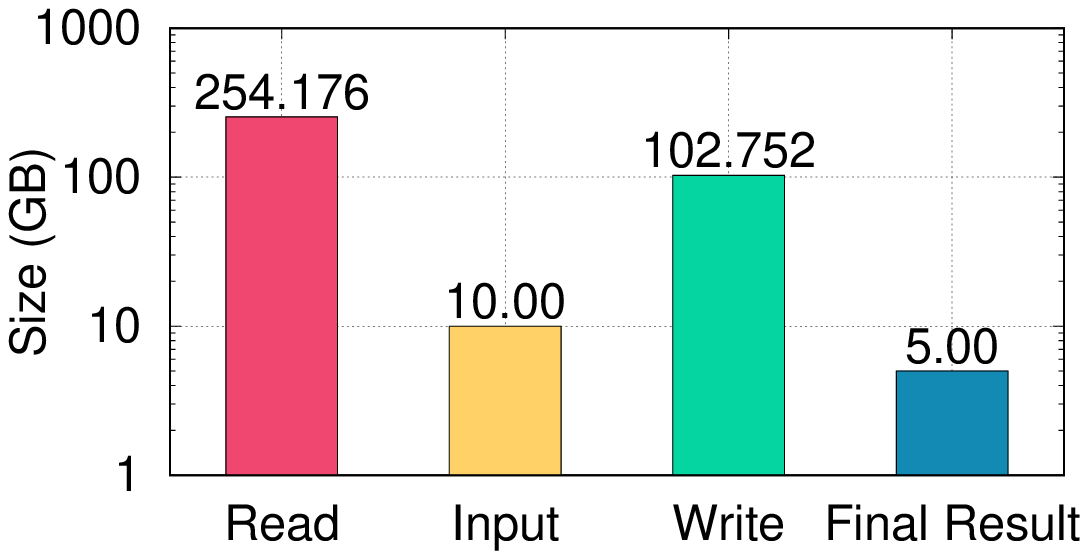}
\end{center}
\vspace{-25pt}
\caption{ 
\small{Numpywren GEMM read and write amplification.}
}
\vspace{-8pt}
\label{fig:moti_gemm}
\end{minipage}
\hfill
\begin{minipage}[b]{0.3\textwidth}
\begin{center}
\includegraphics[width=1\textwidth]{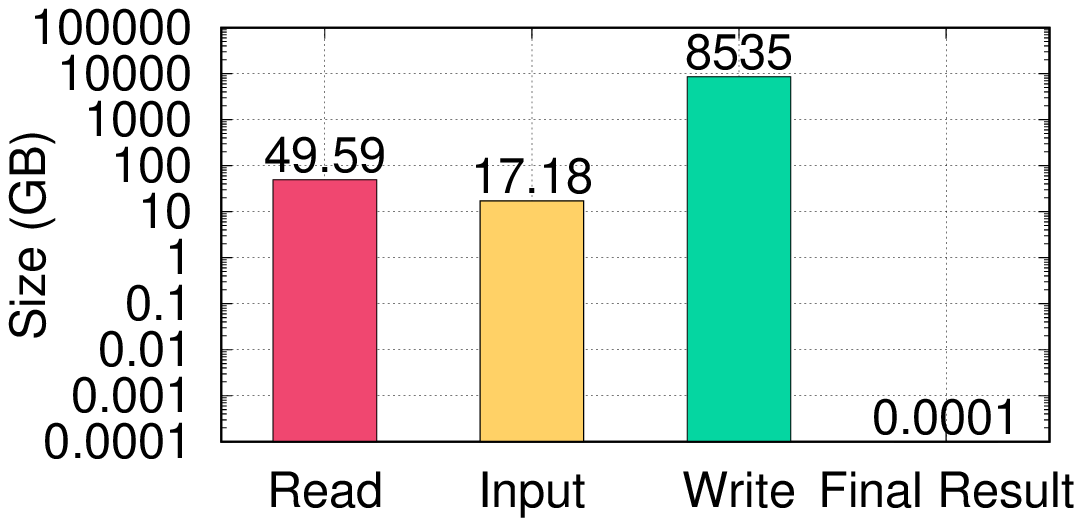}
\end{center}
\vspace{-25pt}
\caption{
\small{Numpywren TSQR read and write amplification. }
}
\vspace{-8pt}
\label{fig:moti_tsqr}
\end{minipage}
\end{minipage}

\end{figure*}

Scheduling is decentralized by partitioning a DAG into multiple, possibly overlapping, subgraphs. Each subgraph is assigned to a task Executor (implemented as an AWS Lambda function runtime) that is responsible for scheduling and executing tasks in its assigned subgraph. This decentralization brings multiple key benefits: 
    
    \noindent\textbf{Enhanced data locality and reduced resource contention:} Decentralization improves the data locality of scheduling. Unlike PyWren~\cite{pywren_socc17} and numpywren~\cite{numpywren}, which require executors to perform network I/Os to obtain each task they execute (since numpywren's task executor is completely stateless), {\proj} preserves task dependency information on the Lambda side. This allows Lambda executors to cache intermediate data and schedule the downstream tasks in their subgraph locally, i.e., without constant remote interaction with a centralized scheduler. 
    
    \noindent{\bf Harnessing scale and local optimization opportunities:} Decentralizing scheduling allows an Executor to make local data-aware scheduling decisions about the level of task granularity (or parallelism) appropriate for its subgraph. Agile executors can scale out compute resources in the face of burst-parallel workloads by partitioning their subgraphs into smaller graphs that are assigned to other executors for an even higher level of parallel task scheduling and execution. Alternately, an executor can execute tasks locally, when the cost of data communication between the tasks outweighs the benefit of parallel execution.
    
    \noindent {\bf Automatic resource elasticity and improved cost effectiveness:} Decentralization does not require users to explicitly tune the number of active Lambdas running as workers and thus is easier to use, more cost effective, and more resource efficient.

We make the following contributions in this paper.
\vspace{-3pt}
\begin{itemize}
    \item We thoroughly explore the problem space of serverless parallel computing framework design. For a range of parallel computation applications, we identify issues of the state-of-the-art serverless frameworks---task scheduling, data locality, and resource efficiency (monetary cost effectiveness).
    \item We present the design and implementation of {\proj}, a new serverless parallel computing framework that solves the identified issues. {\proj} synergizes a set of optimization techniques, including decentralized scheduling, task clustering, and delayed I/O. These techniques together achieve near-ideal scalability, reduce data movement over the network, enhance data locality, and improve cost effectiveness.
    \item We evaluated {\proj} extensively on AWS.
    Our results show that {\proj} reduces network I/O by many orders of magnitude and achieves up to $68.17\times$ higher performance than numpywren,
    while reducing the monetary cost by as much as $92.96\%$.
\end{itemize}

\vspace{-10pt}
\section{Background and Motivation}
\label{sec:moti}

\subsection{Why Serverless?}
\label{subsec:primer}
\vspace{-2pt}

\noindent\textbf{Serverless Computing}
handles virtually all system administration tasks, making it easier for developers to use a near-infinite amount of cloud resources, including 
bundled CPUs and memory, object stores, and a lot more~\cite{serverless_berkeley}. 
Service providers provide a flexible interface for defining serverless functions, which allows developers to focus on core application logic. Service providers in turn auto-scale function executions in a demand-driven fashion, hiding tedious server configuration and management tasks from the users.

\noindent\textbf{General Constraints and Limitations.}
Service providers place limits on the use of cloud resources to simplify resource management. Take AWS Lambda for example: users configure Lambda's memory and CPU resources in a bundle. Users can choose a memory capacity between $128$MB--$3008$MB in $64$MB increments.  Lambda will then allocate CPU power linearly in proportion to the amount of memory configured. Each Lambda function can run at most $900$ seconds and will be forcibly stopped when the time limit is reached. In addition, Lambda only allows outbound TCP network connections and bans inbound connections and the UDP protocol. 

\noindent\textbf{Opportunities.}
Running large-scale, burst-parallel computation jobs 
has long been challenging for domain scientists due to the complexity of configuring, provisioning, and managing compute clusters~\cite{mpi, kubernetes}. 
By taking over system administration and automatically providing capability to launch thousands of processes with no advance notice, the emerging serverless computing model \emph{seems} to provide a foundation that will attract domain scientists and data analysts.
\emph{However, to fully unleash the potential of serverless computing, an efficient serverless-optimized parallel computing framework is needed.}

\vspace{-2pt}
\subsection{Challenges}
\label{subsec:challenges}

We build on our experience with serverless frameworks to synthesize the performance requirement for an ideal, serverless, parallel computing framework, and discuss why current solutions are not able to meet these performance requirements of
burst-parallel applications at both the task scheduling and the data locality level.

\noindent\textbf{Challenge to Rapidly Scale Out.} 
A family of burst-parallel computation jobs (e.g., data analytics~\cite{sparrow_sosp13}, machine learning classifications~\cite{scikit_svm}, etc.) are dominated by short-lived tasks with a span ranging from hundreds of milliseconds (ms) to tens of seconds~\cite{alitrace_iwqos19, sparrow_sosp13, pywren_socc17, psvm_nips07}. 
Such applications pose a difficult scheduling challenge to the serverless computing platforms. This is because, while serverless computing promises to deliver elastic auto-scaling feature in response to bursts of concurrent workloads, serverless function invocations incur non-negligible overhead, thus creating a scheduling bottleneck with slow scaling out. 

PyWren is a state-of-the-art serverless execution engine~\cite{pywren_socc17}. PyWren enables users to program MapReduce-like applications on serverless platforms such as AWS Lambda. 
Numpywren~\cite{numpywren} is a system for linear algebra built on top of PyWren. Numpywren uses PyWren's existing infrastructure to deploy their own serverless task executors, which run as a user-defined function within PyWren's own Lambda executors. In order for numpywren to scale the size of their Lambda cluster, numpywren invokes additional PyWren executors (using PyWren's own API) from their central scheduler. Based on this design, numpywren relies heavily on PyWren for Lambda scaling and management.

Figure~\ref{fig:moti_pywren} shows PyWren's ability to schedule large numbers of no-op tasks on AWS Lambda-based executors. 
Ideally, an omniscient serverless scheduler should be able to take full advantage of the massive parallelism offered by serverless computing and rapidly scale to thousands of Lambda executors in seconds in response to bursty, highly parallel workloads.
PyWren uses a centralized scheduling approach, where it employs 64 threads for task scheduling and invocation; it takes almost 2 minutes to scale out to $10,000$ Lambda executors\footnote{As did in \cite{excamera_nsdi17}, we also performed warmup operations to make sure each Lambda invocation does not incur a cold start~\cite{cold_start_war}.}. 
To make it worse, a serverless framework like PyWren cannot always keep thousands of Lambda executors actively running (unlike the worker servers to a typical serverful parallel framework such as MapReduce~\cite{mapreduce_osdi04}), so it has to constantly invoke many Lambdas in response to bursts of job tasks.

\emph{This serves to illustrate the failure of existing serverless execution frameworks to fully utilize serverless computing's elastic auto-scaling property.}
\noindent\textbf{Challenge of Excessive Data Movement.}
Parallel applications require intermediate data exchange among tasks. Direct task-to-task data communication is naturally supported in traditional serverful parallel computing frameworks such as MPI~\cite{mpi},  MapReduce~\cite{mapreduce_osdi04}, and Dask~\cite{dask}. However, data exchange in serverless applications may be supported only indirectly, through the use of remote cloud storage systems.

Consider the data exchanged during the execution of \(25k \times 25k\) GEMM (general matrix multiplication) and $8,192k \times 128$ TSQR (tall skinny QR) on numpywren.
Figure~\ref{fig:moti_gemm} and Figure~\ref{fig:moti_tsqr} present a comparison between pure input and output sizes and the amount of data transferred during the two workloads respectively. For GEMM, the total quantity of data read is more than \(25\times\) the size of the input data while the total quantity of data written is more than \(20\times\) the size of the output. This trend is further exemplified by TSQR. While the amount of data read is only \(2.88\times\) greater than the input data size, the amount of data written over $65 M\times$ greater than the output size. This is because numpywren and PyWren adopt a stateless Lambda executor design where a task can be dispatched to any Lambda executor; once dispatched to a Lambda, the task simply performs the following four steps: 1) reads its input data (the intermediate data generated by one or multiple upstream tasks) from cloud storage (numpywren uses S3), 2) performs computation,  3) writes the intermediate results (as output) to the cloud storage, and 4) returns. While the stateless design seems to be a good fit for serverless platforms, it does not preserve data locality, which results in excessive data movement. 

\emph{This stresses a strong need for reducing data movement and increasing data locality in serverless frameworks.}
\vspace{-5pt}
\section{{\proj} Design}
\label{sec:design}

In this section, we present the system design of {\proj}. This design is motivated by the observations that existing serverless parallel frameworks are slow to scale out and are bottlenecked by excessive data movement (\cref{sec:moti}).

\vspace{-4pt}
\subsection{High-Level Design}
\label{subsec:HighLevelDesignAndExecution}

\begin{figure}[t]
\begin{center}
\includegraphics[width=0.38\textwidth]{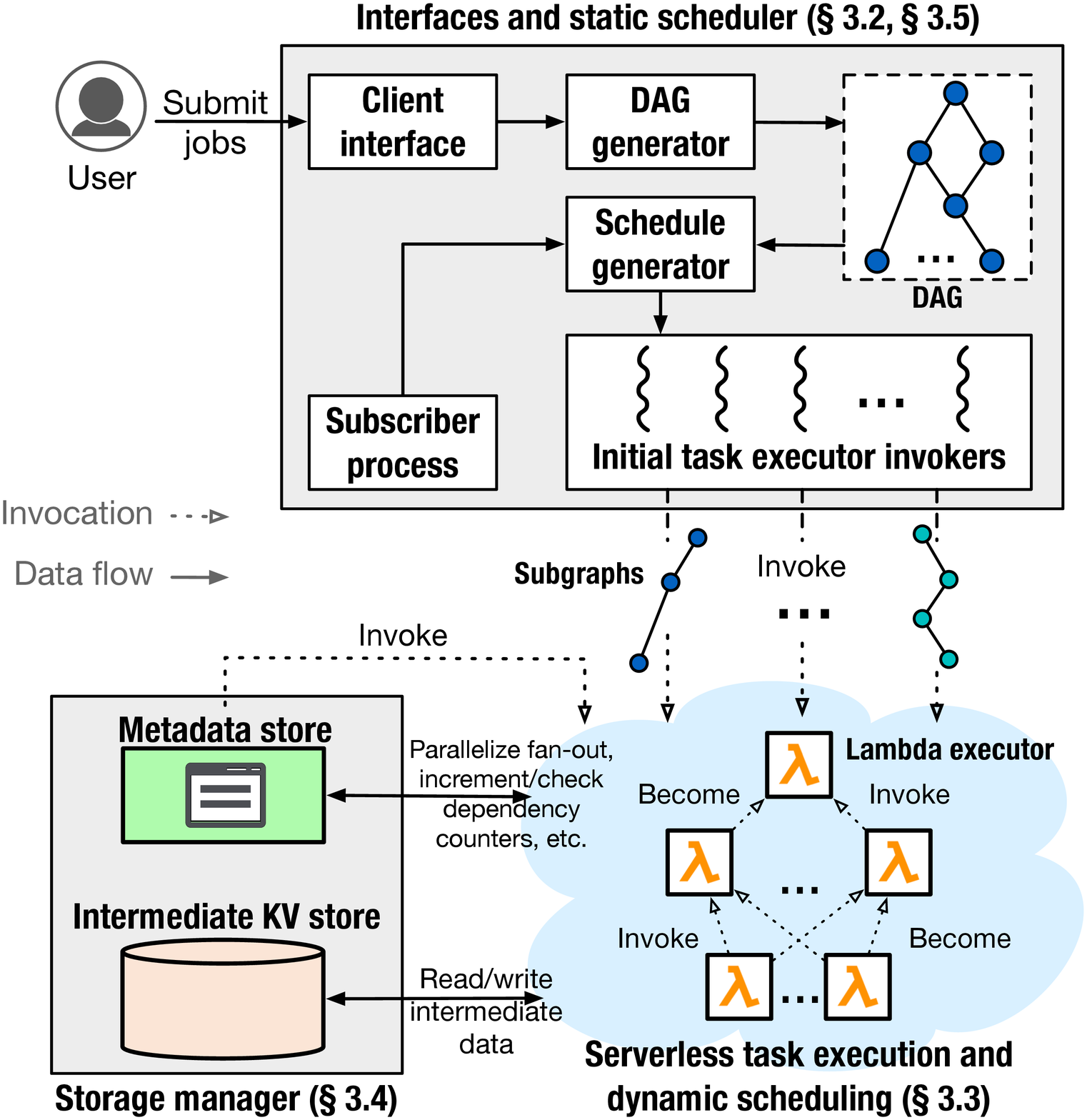}
\vspace{-5pt}
\caption{Overview of {\proj} architecture.}
\vspace{-10pt}
\label{fig:arch}
\end{center}
\end{figure}

Figure~\ref{fig:arch} shows the high-level design of {\proj}. The design consists of three major components: a static schedule generator which runs on Amazon EC2, a pool of Lambda Executors, and a storage cluster.

Scheduling in {\proj} is decentralized and uses a combination of static and dynamic scheduling. A static schedule is a subgraph of the DAG. Each static schedule is assigned to a separate Executor. An Executor uses dynamic scheduling to enforce the data dependencies of the tasks in its static schedule. An Executor can invoke additional Executors to increase task parallelism, or cluster tasks to eliminate any communication delay between them. Executors store intermediate task results in an elastic in-memory key-value storage (KVS) cluster (hosted using AWS Fargate~\cite{aws_fargate}; see \cref{subsec:Storage Management}) and job-related metadata (e.g., counters) in a separate KVS that we call metadata store (MDS).

\vspace{-4pt}
\subsection{Static Scheduling}
\label{subsec:StaticScheduling}

{\proj} users submit a Python computing job to the DAG generator, which uses the Dask library to convert the job into a DAG. The Static-Schedule Generator generates \emph{static schedules} from the DAG. For a DAG with $n$ leaf nodes, $n$ static schedules are generated. A static schedule for leaf node {\small\texttt{L}} contains all of the task nodes that are reachable from {\small\texttt{L}} and all of the edges into and out of these nodes. The data for a task node includes the task's code and the KVS keys for the task's input data. The schedule for {\small\texttt{L}} is easily computed using a depth-first search (DFS) that starts at {\small\texttt{L}}. Lambda Executors notify the static scheduler when a final result has been stored in Redis by sending a message to the static scheduler's subscribe process. Upon receiving a message, the static scheduler will download final results and return them to the user automatically.

\begin{figure}[t]
\begin{center}
\includegraphics[width=0.4\textwidth]{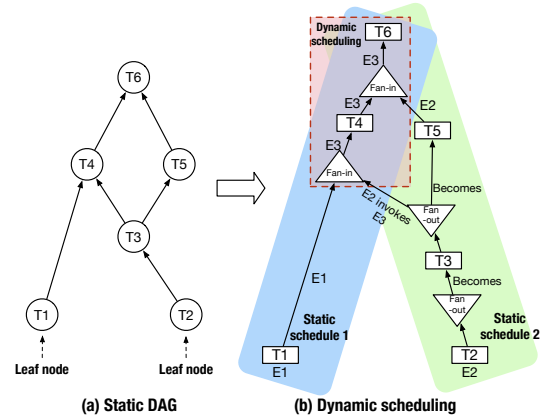}
\vspace{-5pt}
\caption{Static DAG (a) and dynamic scheduling (b). 
{\proj}'s Executors coordinate in the area inside the dashed box in (b) using dynamic scheduling. ``T1'' denotes Task 1. ``E1'' denotes Lambda Executor 1.
}
\vspace{-15pt}
\label{fig:sched}
\end{center}
\end{figure}

Figure~\ref{fig:sched}(a) shows a DAG with two leaf nodes. Figure~\ref{fig:sched}(b) shows the two static schedules that are generated from the DAG: {\small\texttt{Schedule 1}} (blue) and {\small\texttt{Schedule 2}} (green).

A static schedule contains three types of operations: task execution, fan-in and fan-out.
To simplify our description, when DAG task {\small\texttt{Tx}} is followed immediately by task {\small\texttt{Ty}}, and {\small\texttt{Tx}} {\small\texttt{(Ty)}} has no fan-out (fan-in), we add a trivial fan-out operation between {\small\texttt{Tx}} and {\small\texttt{Ty}} in the static schedule. This fan-out operation has one incoming edge from {\small\texttt{Tx}} and one outgoing edge to {\small\texttt{Ty}}, i.e., there is no actual fan-out. In Figures~\ref{fig:sched}(a) and (b), this is the case for DAG tasks {\small\texttt{T2}} and {\small\texttt{T3}}.

A fan-in task {\small\texttt{T}} may depend on tasks that will be executed by different Executors, e.g., task {\small\texttt{T4}} in Figure~\ref{fig:sched}. The dynamic scheduling technique described below ensures that {\small\texttt{T}}'s data dependencies are satisfied and that {\small\texttt{T}} is executed by only one Executor. Note also that a static schedule does not map its tasks to processors; this mapping is done automatically by the AWS Lambda platform when an Executor function instance is invoked with the static schedule and placed on a VM by AWS Lambda.

\vspace{-6pt}
\subsection{Task Execution \& Dynamic Scheduling}

\noindent\textbf{Task Execution.}
{\proj} workflow execution starts when the static scheduler's Initial-Executor Invokers assign each static schedule produced by the Static-Schedule Generator to a separate Executor. Recall that each static schedule begins with one of the leaf node tasks in the DAG. The Initial-Executor invokes these "leaf node" Executors in parallel. After executing its leaf node task, each Executor then executes the tasks along a single path through its schedule. 
An Executor may execute a sequence of tasks before it reaches a fan-out operation with more than one out edge or it reaches a fan-in operation. \emph{For such a sequence of tasks, there is no communication required for making the output of the earlier tasks available to later tasks for input.} All intermediate task outputs are cached in the Executor's local memory with inherently enhanced data locality. Executors also look ahead to see what data their tasks may need, which allows them to discard data in their local memory that is no longer needed.

Furthermore, Executors can increase parallelism by scheduling and invoking new Executors to execute the task targets of a fan-out. A group of Executors that reach a common fan-in node decrease parallelism by scheduling one of them to execute the fan-in task and the rest to stop. {\proj} uses dynamic scheduling to resolve conflict on the fly. More importantly, \emph{this dynamic scheduling of the tasks in an Executor's static schedule leads to a decentralized scheduling model that naturally fits serverless computing, eliminates the need for a centralized scheduler processes that check data dependencies and invoke ready tasks with improved scalability.} 

\noindent\textbf{Dynamic Scheduling for Fan-Out Operations.}
For fan-out operations there are two cases:

\noindent\textbf{Case 1:} none of the $n$ (where $n > 1$) fan-out edges is a fan-in edge.  Then {\small\texttt{E}} ``becomes'' the Executor for one of the fan-out's tasks, say {\small\texttt{T}}, by executing {\small\texttt{T}}, and {\small\texttt{E}} ``invokes'' an Executor for the other fan-out tasks.

\noindent\textbf{Case 2:} one or more of the fan-out edges is also a fan-in edge. For example, fan-out node 3 in Figure~\ref{fig:sched}(a) has a fan-out edge that is a fan-in edge to node 4. The selection of a ``becomes'' edge for {\small\texttt{E}} is based on the immediate availability of the tasks targeted by the fan-out edges.
If no task target's dependencies are satisfied then no task target is immediately available for execution and none of the fan-out edges can be selected as {\small\texttt{E}}'s ``becomes'' edge (the fan-in edges have fan-in operations that will be executed next); otherwise, one of the fan-out edges for the available target tasks is selected as the ``becomes'' edge.

An intermediate objects needed for input by an invoked Executor is passed to the Executor as an argument if the size of the object is less than the maximum allowed argument size (256K);otherwise, the object is sent to the Storage Manager, and the associated KVS keys are passed to the invoked Executors as arguments. 

Each of the $n-1$ Executors invoked by {\small\texttt{E}} is assigned a static schedule that begins with one of the $n-1$ fan-out edges. Each of these (possibly overlapping) static schedules corresponds to a sub-graph of {\small\texttt{E}}'s static schedule. Executor {\small\texttt{E}} continues task execution and scheduling along the remaining fan-out edge and executes the operation encountered on this edge. 

In Figure~\ref{fig:sched}(b), fan-out edges are labeled either ``invokes'' or ``becomes'' to indicate whether the Executor invokes a new Lambda Executor to execute a fan-out task or executes the fan-out task itself.

Since Executor invocations, which are in the form of AWS Lambda function invocations, incur a high overhead (e.g., invoking an AWS Lambda function takes about 50 milliseconds with the Boto3 AWS Python API), we use a number of dedicated Executor-Invoker processes that are co-located with the Static Scheduler (Figure~\ref{fig:arch}). When an Executor performs a fan-out operation, and a large number of new Executors must be invoked, the Executor delegates the invocations to the Static Scheduler. The Static Scheduler evenly distributes task invocation responsibilities among the Invoker processes, enabling (near-)linear speedup over sequential invocations.

\noindent\textbf{Dynamic Scheduling for Fan-in Operations.} 
If Task Executor {\small\texttt{E}} executes a fan-in operation with $n$  (where $n > 1$) in-edges, then {\small\texttt{E}} and the $n-1$ other Executors involved in this fan-in operation cooperate to see which one of them will continue their static schedules on the out edge of the fan-in (e.g., node 4 in Figure~\ref{fig:sched}).

For a fan-in operation executed by {\small\texttt{E}} for fan-in task {\small\texttt{T}}, {\small\texttt{E}} atomically gets and updates a value in the KVS that tracks the number of {\small\texttt{T}}'s input dependencies that have been satisfied during execution
There are two cases:

\noindent\textbf{Case 1:} all of the input dependencies of {\small\texttt{T}} have been satisfied. Then E continues its static schedule by executing {\small\texttt{T}}

\noindent\textbf{Case 2:} all of the input dependencies of {\small\texttt{T}} have not been satisfied. Then {\small\texttt{E}} sends the intermediate object needed by {\small\texttt{T}} to the Storage Manager. 

In Figure~\ref{fig:sched}(b), each fan-in task is labeled with the Executor that executed the task.

\noindent\textbf{Task Clustering.} Storing and retrieving large intermediate objects can be very costly. {\proj} implements task clustering to avoid large object storage.

\noindent\textbf{Task Clustering for Fan-Out Operations.}
If the output object of some fan-out task \(T\) executed by Executor {\small\texttt{E}} is larger than a user-defined threshold \(t\) (e.g., 200~MB), {\small\texttt{E}} will try to cluster the target tasks of {\small\texttt{T}}'s fan-out edges, i.e.,  {\small\texttt{E}} will execute the target tasks whose dependencies are satisfied instead of invoking new Executors for these tasks. For example, 
the  Executor that executes task C can also execute tasks F and G to avoid the time and cost of communicating task C's large object output to tasks F and G. In cases like this, the fan-out edges will have multiple edges labeled ``becomes''.

\noindent\textbf{Task Clustering for Fan-In Operations.} If {\small\texttt{E}} executes a fan-in operation for task {\small\texttt{T}} and the input dependencies of a single task {\small\texttt{T}} have not been satisfied, then {\small\texttt{E}} sends the intermediate object needed by {\small\texttt{T}} to the Storage Manager. If this object is large, {\small\texttt{E}} then rechecks {\small\texttt{T}}'s input dependencies. If {\small\texttt{T}}'s input dependencies became satisfied while the large intermediate object was being stored, {\small\texttt{E}} becomes the Executor for {\small\texttt{T}}, This avoids the communication delay that would have occurred when {\small\texttt{T}} retrieved the large object from storage, but not the delay that occurs when {\small\texttt{R}} stored the object. The delay that occurred when the large object was stored by {\small\texttt{E}} can be avoided if the storage operation can be delayed until after {\small\texttt{T}}'s input dependencies become satisfied. 

Suppose that when Executor {\small\texttt{E}} executes the fan-out operation it identifies multiple fan-out tasks that are also fan-in tasks and that have input dependencies that are not satisfied.  {\small\texttt{E}} will then execute any ready fan-out tasks it finds and delay the decision about how to handle the fan-out tasks that are not ready. (There may be many fan-out tasks that are not also fan-in tasks or that are fan-in tasks with satisfied input dependencies.) 

\noindent\textbf{Delayed I/O.}
After Executor {\small\texttt{E}} executes the ready tasks,  {\small\texttt{E}} will recheck whether the input dependencies of the unready tasks have since become satisfied. If so, the newly ready tasks can be executed. If at least one unready task becomes ready, the ready task can be executed and the unready tasks can  be rechecked again, and so on, for a configurable number of times. Our profiling indicates that it is almost always better to wait until all of the unready tasks become ready to avoid the very large communication delay the occurs when many large objects are stored and retrieved potentially at the same time. A pattern of simultaneous large object writes and reads occurred often in the DAG workloads that we executed.  If unready tasks remain when this process stops, {\small\texttt{E}} must send the intermediate objects needed by the unready tasks to storage; however, as described above, {\small\texttt{E}} can possibly avoid the communication delay associated with retrieving the objects from storage by checking the unready objects one more time.

\vspace{-10pt}
\subsection{Storage Management}
\label{subsec:Storage Management}

{\proj}'s Storage Cluster is built atop AWS Fargate~\cite{aws_fargate}, a serverless container engine that can be elastically scaled out/in. The Storage Cluster includes a number of AWS Fargate tasks, each of which hosts a Redis instance, and a KV Store Proxy Service. The KV Store Proxy is executed within an EC2 VM along with an additional instance of Redis used exclusively for storing static schedules and dependency counters. The user can configure the size of the Fargate cluster to dynamically accommodate workloads of different sizes. The user simply specifies how many nodes they would like, and {\proj} ensures that these nodes are created and available. The Fargate nodes are used for the storage of intermediate data generated during a workload's execution. 
We opt to use Redis instead of S3 as Redis can provide both high bandwidth for large object workloads and high IOPS for small object workloads~\cite{pocket_osdi18, locus_nsdi19}, whereas S3's IOPS is throttled. However, modifying {\proj} to use S3 is trivial.

The KV Store Proxy performs various storage operations on behalf of the Task Executors and the Scheduler. At the start of a workflow, the Storage Manager receives the workflow DAG and the static schedules derived from the DAG from the Scheduler.

\noindent\textbf{Intermediate and Final Result Storage.} Task Executors publish their intermediate and final task output objects to the KV Store. Final outputs are relayed to a Subscriber process in the Scheduler for presentation to the Client.

\noindent\textbf{Small Fan-out Task Invocations.} When a Task Executor performs a fan-out operation that has a small number of out edges, the Task Executor makes the necessary Executor invocations itself. 

\noindent\textbf{Large Fan-out Task Invocations.} When a fan-out has a number of out edges that is larger than a user-specified threshold, the Task Executor publishes a message that is relayed to a Subscriber process in the Storage Manager that then passes the message on to the Proxy. This message contains an ID that identifies the fan-out's location in the DAG. The Proxy uses the DAG and the fan-out ID to identify the fan-out's out edges in the DAG. This allows the Proxy, with the assistance of the Fan-out Invokers in the Storage Manager, to make the necessary Task Executor invocations, in parallel. The Proxy passes to each Executor its intermediate inputs (or their key values in the KV Store) and the Executor's static schedule.

\begin{figure}
\vspace{-10pt}
\begin{minipage}{\textwidth}

\begin{minipage}[b]{0.34\textwidth}
\begin{center}
\begin{minted}[fontsize=\small, xleftmargin=2pt, numbersep=2pt,linenos]{python}
  def add(x,y):
    time.sleep(0.5)
    return x + y

  L = range(8)
  while len(L) > 1:
    L = list(map(delayed(add),\ 
            L[0::2], L[1::2]))

  L[0].compute()
    \end{minted}
\end{center}
\vspace{-12pt}
\caption{
\small{TR code.}}
\label{fig:tr_code}
\end{minipage}
\hspace{-4em}%
\begin{minipage}[b]{0.26\textwidth}
\begin{center}
\includegraphics[width=0.6\textwidth]{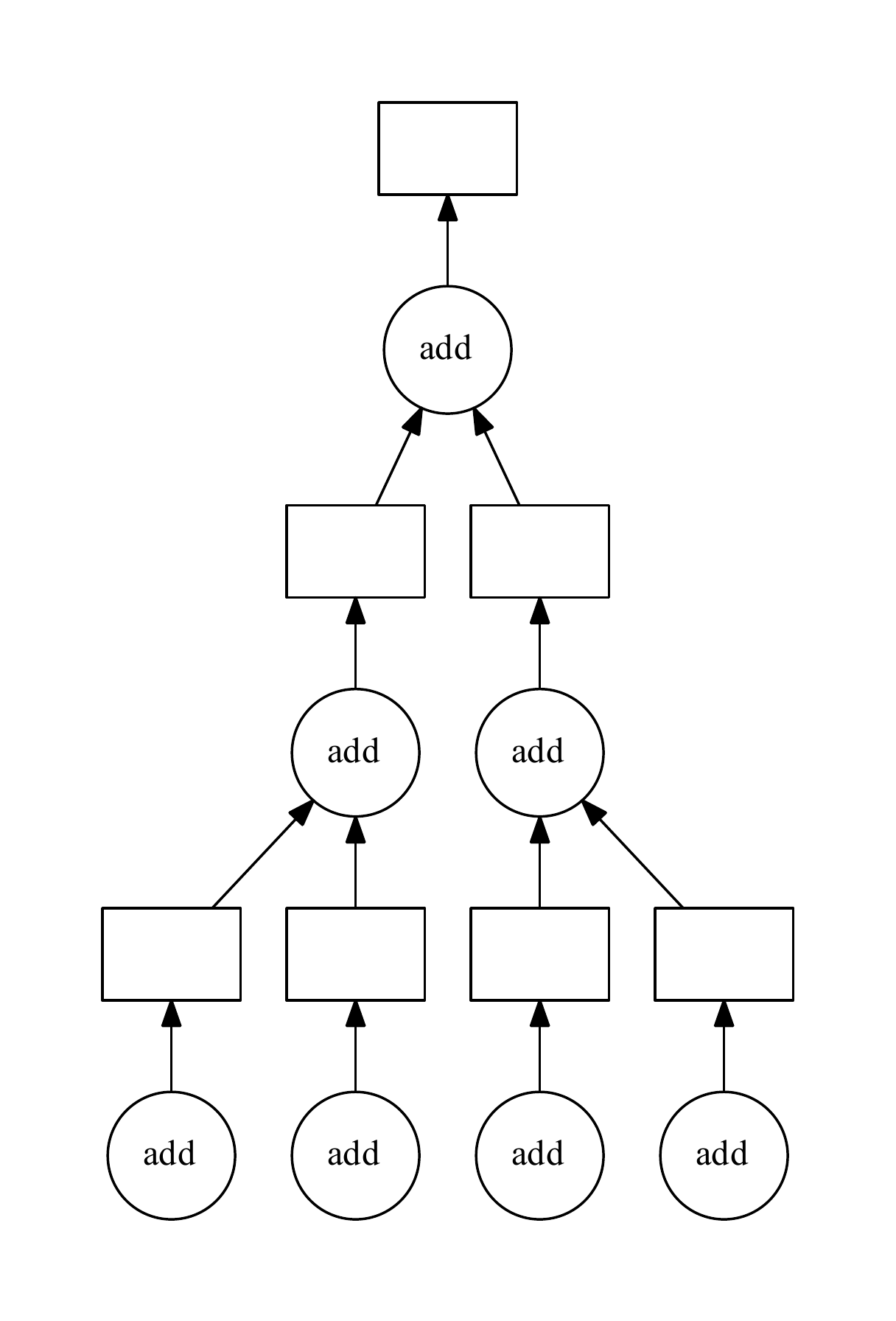}
\end{center}
\vspace{-12pt}
\caption{
\small{TR DAG.}}
\label{fig:tr_dag}
\end{minipage}

\end{minipage}
\vspace{-20pt}
\end{figure}

\vspace{-10pt}
\subsection{Programming Model}

{\proj} reuses Dask's Python programming interfaces~\cite{dask_api} (including high-level APIs, such as Dask libraries and data structures, and low-level APIs, such as {\small\texttt{Dask.delayed}}) for implementing parallel programs. In general, any code written for Dask should execute on {\proj}. Figure~\ref{fig:tr_code} shows an example code snippet that implements tree reduction (TR) (a detailed description of TR is in \cref{subsec:methodology}). 
{\proj} also reuses Dask's DAG generator which translates high-level Python code into a DAG~\cite{dask_dag}. Figure~\ref{fig:tr_dag} depicts the DAG generated for TR with an 8-element array.

\vspace{-6pt}
\subsection{Fault Tolerance}

For fault tolerance, we relied on the automatic retry mechanism of AWS Lambda, which allows for up to two automatic retries of failed function executions. Investigating better fault tolerance scheme is part of our future work.
\vspace{-6pt}
\section{Evaluation}
\label{sec:eval}

\noindent\textbf{Implementation.} We have implemented {\proj} using roughly $12k$ lines of Python code ($5,349$ LoC for the AWS Lambda Executor Runtime, $3,057$ LoC for the Storage Manager, and $3,577$ LoC for the Static Scheduler). We use the Dask library~\cite{dask} to generate DAGs to use as the input computation graph for {\proj}.

\vspace{-12pt}
\subsection{Experimental Goals and Methodology}
\label{subsec:methodology}

Our evaluation was performed on AWS. The static scheduler ran in an {\small{\texttt{r5n.16xlarge}}} EC2 VM. We used a scale-out Redis cluster (Multi-Redis) as {\proj}'s intermediate storage. The Fargate nodes within the storage cluster were each configured to have 30~GB of memory and 4 vCPUs. 
For the Multi-Redis setup, {\proj} used 75 nodes for the storage cluster, as we've determined this value to be both performant and cost-effective for many workloads.
The MDS proxy was co-located on the same {\small{\texttt{r5n.16xlarge}}} VM as a Redis instance that was used for storing static schedules and dependency counters. Each Lambda function was allocated 3~GB of memory and a maximum execution time of seven minutes.

We compared the performance of {\proj} against both Dask distributed (which we refer to simply as ``Dask'') and numpywren~\cite{numpywren} / PyWren~\cite{pywren_socc17}. We chose to compare our performance against Dask as both {\proj} and Dask use the exact same input DAG and the same algorithms for their computations. We compared the end-to-end performance of {\proj} against numpywren, and we compared the scalability of {\proj} against PyWren, which is numpywren's underlying Lambda execution framework.

We chose to compare {\proj} against numpywren because both are serverless DAG execution frameworks; however, there are significant differences between {\proj} and numpywren. One difference is that {\proj} uses Dask DAGs, which explicitly encode tasks and their dependencies. Numpywren, on the other hand, uses an implicit DAG representation for its programs, which are all implemented in the LAmbdaPACK language for linear algebra~\cite{numpywren}. The nodes of the DAG that represent a LAmbdaPACK program are generated on-demand (at runtime). 

Another difference is that numpywren uses AWS S3 as its intermediate data store. In order to make the comparison between {\proj} and numpywren more fair, we modified numpywren to use a single instance of Redis as its intermediate object store. We compared this version of numpywren, which we refer to as ``Numpywren Single Redis'', with a modified version of {\proj} that also uses a single instance of Redis for intermediate data storage. In addition, we compared numpywren S3 against ``{\proj} Multi-Redis'', which uses a Fargate Redis cluster for its intermediate object store. This second comparison was performed in order to show the optimal performance achieved by each system.

Finally, numpywren only supports linear algebra algorithms and only several of these algorithms are also available in Dask. As a result, we are limited in which workloads we can run on both {\proj} and numpywren.

To better understand {\proj}'s performance, we compared {\proj} against 2 different Dask configurations. Both configurations used the same amount of CPU ($2,000$ cores) and memory ($3,000$~GB memory). 
This was the largest VM cluster that we could configure. 
The first configuration consisted of $1,000$ 2-core 3GB workers running on 125 {\small\texttt{c5.4xlarge}} 16-core 32~GB VMs. Each VM had eight workers running on it. The second configuration consisted of 125 workers; each worker exclusively ran on a {\small\texttt{c5.4xlarge}} VM and was allocated 16 cores and 24~GB memory. 
The first configuration was selected so that each worker was approximately as powerful as the AWS Lambda functions used by {\proj}; it represents a \emph{worst-case scenario} that stresses the Dask scheduler with many workers and incurs high communications due to the lack of data locality, emulating a serverless environment with centralized scheduling. The second configuration represents a \emph{best-case scenario} where the relatively more powerful workers could process larger data with improved data locality and significantly reduced communications. 

Additionally, we used the exact same input data partitions for Dask and {\proj}. We scaled the problem size for each benchmark by having each (Dask, {\proj}, or numpywren) worker assigned with (at least) one partition of the input data. As such, a  test used only a fraction of the $2,000$ cores of the Dask cluster, until the problem size was large enough to occupy all the resources.
The largest number of partitions (of input data) used during our evaluation was $4,096$ for  $16~M$ TSQR  (which will be described shortly). This number of partitions does not overwhelm either Dask configurations as TSQR's DAG features large fan-ins in the middle of the job. The large fan-ins result in the number of tasks allocated to each worker decreasing as the workload progresses.  

\begin{figure*}
\vspace{-15pt}
\begin{minipage}{\textwidth}
\begin{minipage}[b]{0.26\textwidth}
\begin{center}
\includegraphics[width=1\textwidth]{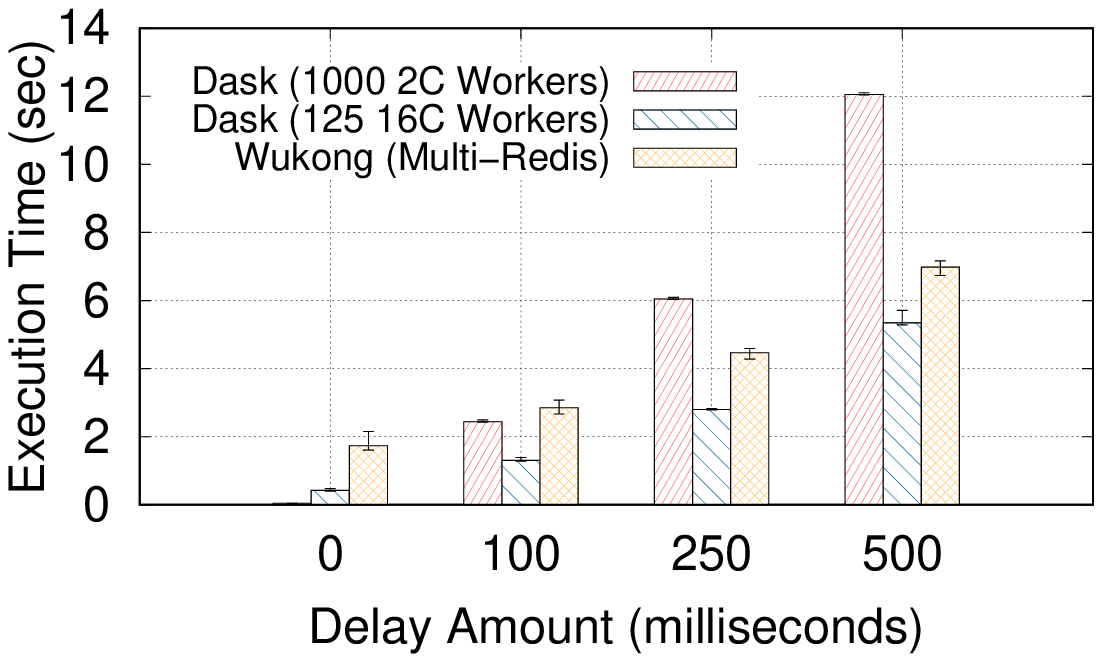}
\end{center}
\vspace{-20pt}
\caption{
\small{TR.}
}
\label{fig:tr-eval}
\end{minipage}
\hfill
\hspace{-3em}%
\begin{minipage}[b]{0.26\textwidth}
\begin{center}
\includegraphics[width=1\textwidth]{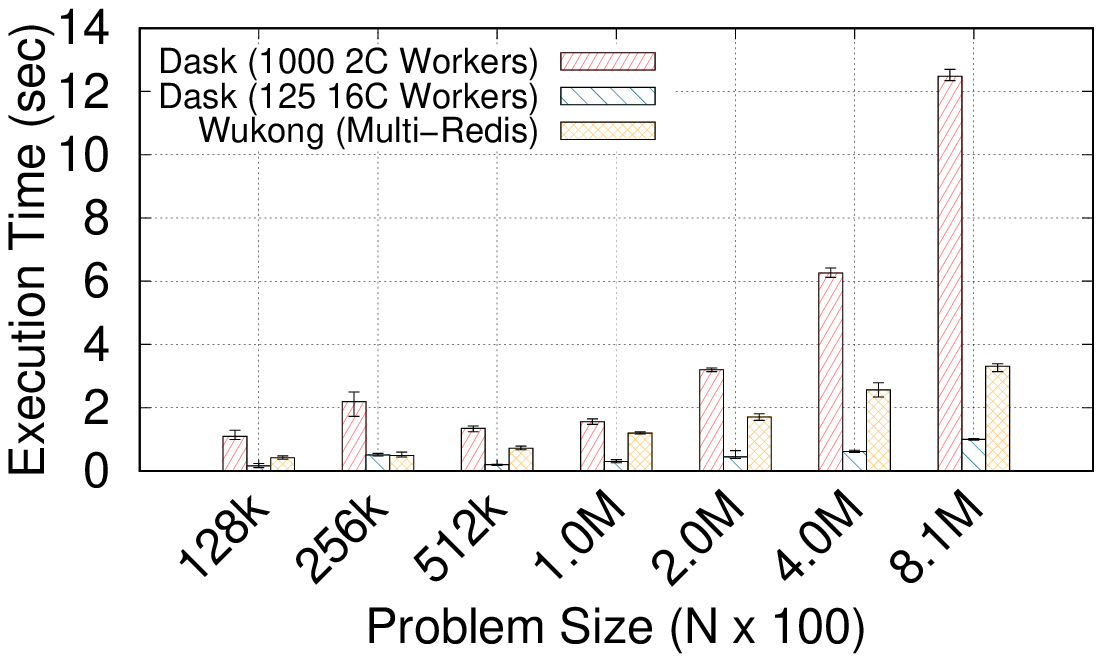}
\end{center}
\vspace{-20pt}
\caption{ 
\small{SVD1.}
}
\label{fig:svd1-eval}
\end{minipage}
\hfill
\hspace{-3em}%
\begin{minipage}[b]{0.26\textwidth}
\begin{center}
\includegraphics[width=1\textwidth]{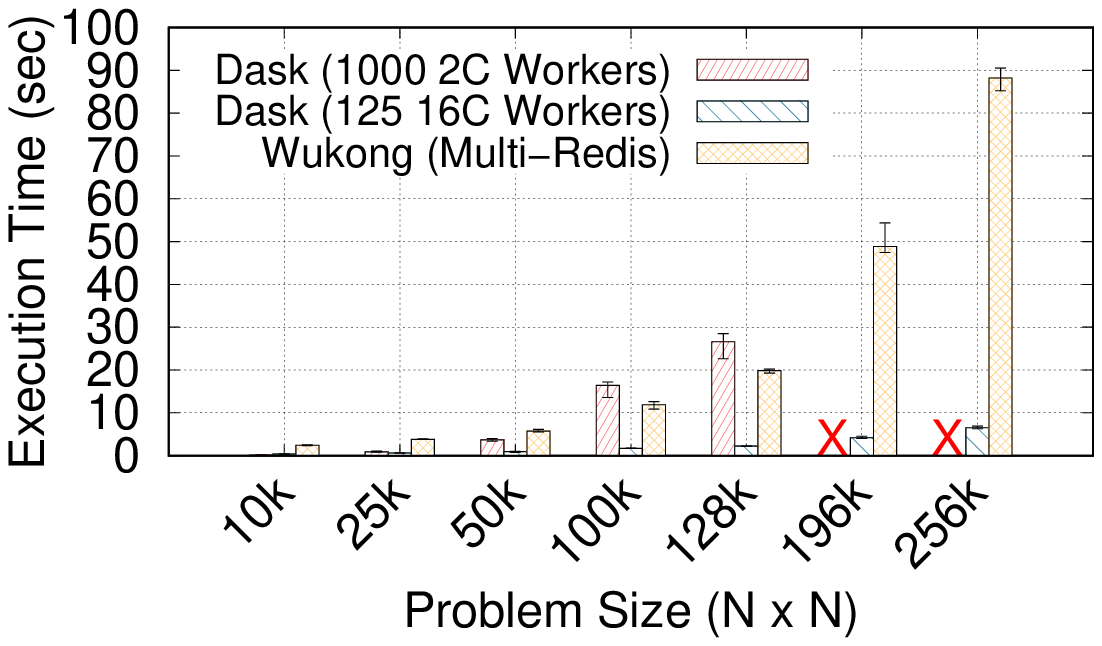}
\end{center}
\vspace{-20pt}
\caption{
\small{SVD2. }}
\label{fig:svd2-eval}
\end{minipage}
\hfill
\hspace{-3em}%
\begin{minipage}[b]{0.26\textwidth}
\begin{center}
\includegraphics[width=1\textwidth]{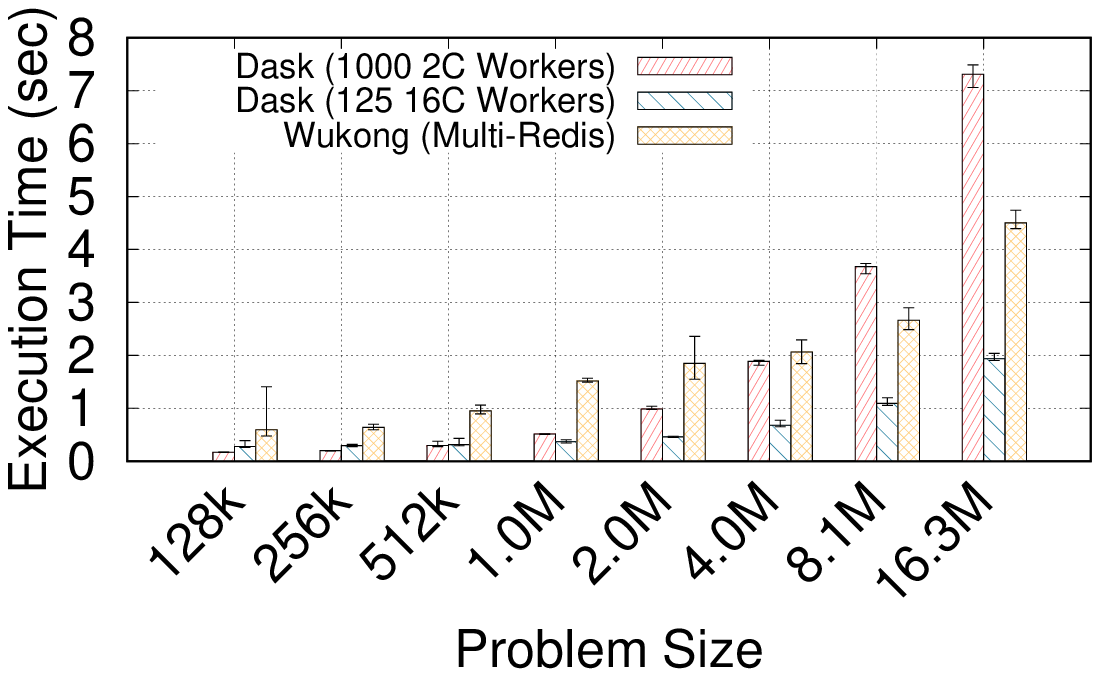}
\end{center}
\vspace{-20pt}
\caption{ 
\small{SVC.}}
\label{fig:svc-eval}
\end{minipage}

\end{minipage}

\vspace{-10pt}
\end{figure*}

\begin{figure*}
\vspace{-10pt}
\begin{minipage}{\textwidth}

\begin{minipage}[b]{0.26\textwidth}
\begin{center}
\includegraphics[width=1\textwidth]{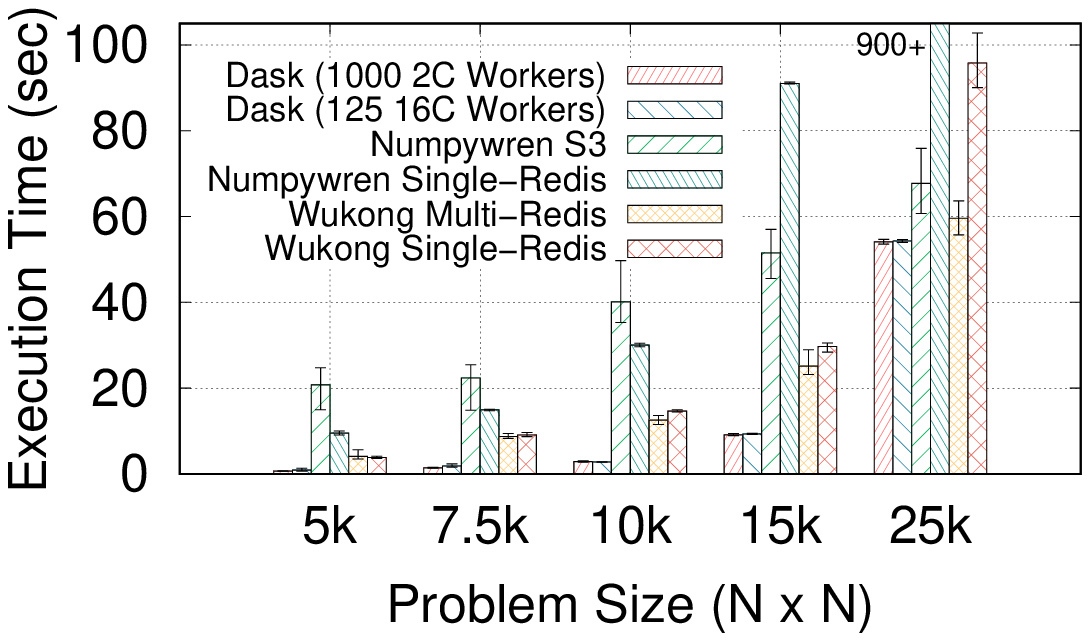}
\end{center}
\vspace{-20pt}
\caption{
\small{GEMM.}}
\label{fig:gemm-eval}
\end{minipage}
\hfill
\hspace{-3em}%
\begin{minipage}[b]{0.26\textwidth}
\begin{center}
\includegraphics[width=1\textwidth]{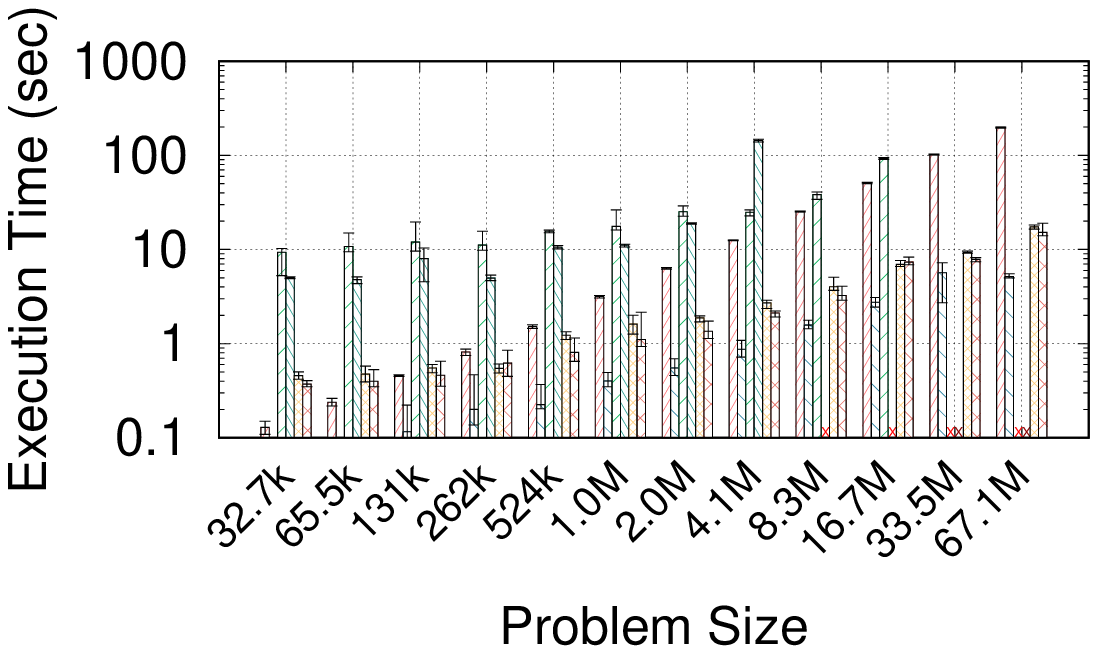}
\end{center}
\vspace{-20pt}
\caption{
\small{TSQR (log-scale).}}
\label{fig:tsqr-eval}
\end{minipage}
\hfill
\hspace{-3em}%
\begin{minipage}[b]{0.26\textwidth}
\begin{center}
\includegraphics[width=1\textwidth]{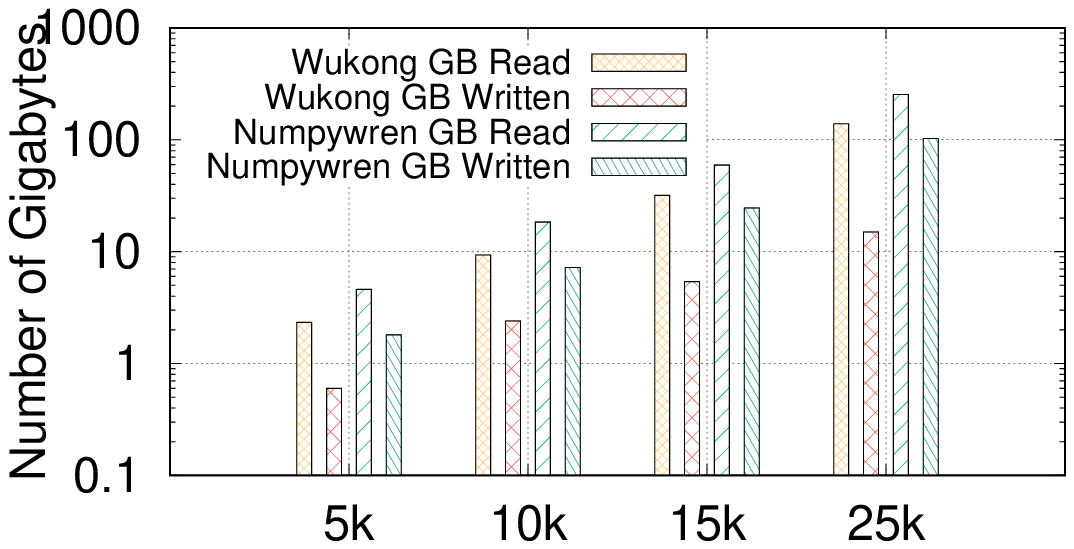}
\end{center}
\vspace{-20pt}
\caption{
\small{GEMM I/O (log).}}
\label{fig:gemm-io}
\end{minipage}
\hfill
\hspace{-3em}%
\begin{minipage}[b]{0.26\textwidth}
\begin{center}
\includegraphics[width=1\textwidth]{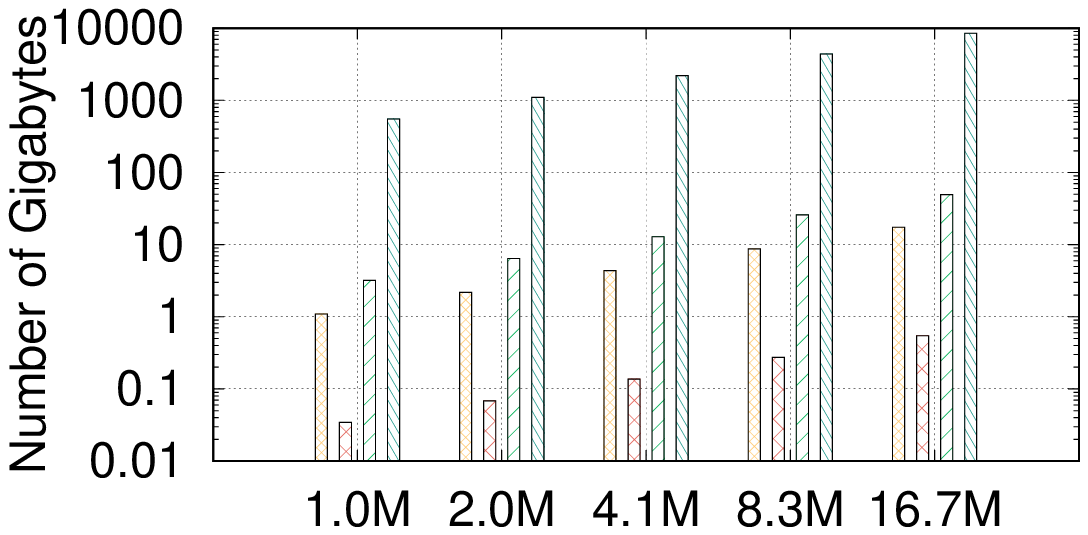}
\end{center}
\vspace{-20pt}
\caption{
\small{TSQR I/O (log).}}
\label{fig:tsqr-io}
\end{minipage}

\end{minipage}
\vspace{-10pt}
\end{figure*}

In our evaluation results, each data point is the average of \emph{ten} runs. The error bars on the graphs of the results indicate the biggest and smallest results obtained. {\proj} is easy-to-use as it exposes only two configuration knobs to the end users---the input partition size and the number of Fargate nodes. (A sensitivity analysis of these two configuration knobs is omitted due to space constraint.)

We evaluated the following parallel applications. 

\noindent\textbf{Tree Reduction (TR):} TR sums the N elements of an array using a total of $N-1$ operations performed over multiple passes. Each pass adds adjacent elements, in parallel, until after $log(N)$ passes only a single element remains. TR serves as a microbenchmark for evaluating the effect of task granularity and parallelism on performance. See Figure~\ref{fig:tr_code} and Figure~\ref{fig:tr_dag} for an example of the Python code snippet and generated DAG. 

\noindent\textbf{Singular Value Decomposition (SVD):} We evaluated two variants of SVD. The first variant (SVD1) computes the SVD of a tall skinny matrix.
The second variant (SVD2) computes the SVD of a square (i.e., \(n \times n\)) matrix using an approximation algorithm provided by ~\cite{halko_finding_2009}. Note that Dask's SVD algorithm differs considerably from numpywren's, and thus a direct comparison between {\proj} and numpywren for SVD is impossible. 
For reference, {\proj} can execute SVD $256k \times 256k$ in 88 seconds on average while \cite{numpywren} reports an average of 77,828 seconds for the same problem size for SVD.

\noindent\textbf{Support Vector Classification (SVC):} SVC is a  machine learning workload. 
The benchmark we used is publicly available in the Dask-ML documentation~\cite{dask-ml}. 

\noindent\textbf{General Matrix Multiplication (GEMM):} GEMM performs matrix multiplication, an important component of many linear algebra algorithms. 

\noindent\textbf{Tall-and-Skinny QR Factorization (TSQR):} This workload performs a QR factorization of a tall skinny matrix.

\vspace{-10pt}
\subsection{End-to-End Performance Comparison}
\label{subsec:e2e}

\noindent\textbf{TR.} The size of the array used for TR was 1024. We also intentionally added a delay to each task of TR. 
This delay simulates an increased amount of work performed per task. 
We varied the amount of delay between 0--500~ms. 
Figure \ref{fig:tr-eval} shows that both configurations of Dask outperforms {\proj} by a large margin for the base case. This is because Dask uses TCP to dispatch the $1024 / 2 = 512$ {\small\texttt{addition}} tasks to workers, whereas for {\proj} the overhead of scaling out to 512 Lambda executors dominates.
As we add increasing amounts of delay to each task, the performance gap between Dask and {\proj} decrease. Once the delay is 250~ms or more, {\proj} executes the workload faster than the $1,000$-worker Dask cluster, as {\proj} uses decentralized scheduling to rapidly scale out to 512 workers. This experiment shows an interesting tradeoff between per-task execution time and serverless scaling cost -- {\proj} performs best when each task performs a non-trivial amount of work, as this compensates for the overhead of spinning up additional Lambdas. 

\noindent\textbf{SVD.} 
We tested seven problem sizes for both SVD1 and SVD2
(see Figure~\ref{fig:svd1-eval} and Figure~\ref{fig:svd2-eval}). 
For nearly all problem sizes, {\proj} out-performed the $1,000$-worker Dask cluster, completing the job anywhere from 62.02\% to 69.09\% faster than Dask. This performance difference results from the benefits of {\proj}'s decentralized scheduling techniques, which greatly reduce the overhead of executing tasks on a large number of workers. The $1,000$-worker Dask was bottlenecked by its central scheduler due to the increasing load from the one thousand workers. 

The 125-worker Dask cluster consistently outperformed {\proj}. This is because each Dask worker is provisioned with more resources (i.e., more CPUs, greater network bandwidth, a larger amount of RAM per worker, etc.), which significantly increases data locality and reduces cross-worker communications.
More importantly, for SVD2, {\proj} was able to scale to considerably larger problem sizes than what the 1,000-worker Dask cluster
was capable of handling (crosses in Figure~\ref{fig:svd2-eval}). 
The results demonstrate \proj's ability to provide competitive performance with traditional serverful  frameworks, while also scaling effectively for increasingly large problem sizes. {\proj}'s ability to scale effectively here is largely because of its use of task clustering and delayed I/O. These techniques dramatically reduce data movement and ensure all downstream tasks which depend on large data are executed on the Task Executor that already has the data in-memory. A quantitative analysis of the benefits of these techniques is given later in \cref{subsec:factor}.

\noindent\textbf{SVC.} Figure~\ref{fig:svc-eval} shows  SVC's performance. 
As with the previous benchmark, Dask was able to perform better for smaller problem sizes; however, when we increased the problem size, the performance gap between the two frameworks decreased. 
As the scale of the problems increased to $4.0M$ samples and beyond, {\proj} began to scale more effectively than the $1,000$-worker Dask cluster. The large parallelism of {\proj} enabled the framework to complete the workload in 46.45\% less time; however, the $125$-worker Dask cluster executed the workload roughly 50.56\% faster than {\proj}. This is likely due to both the higher network bandwidth and increased data locality of the Dask workers.

\noindent\textbf{GEMM.} Like TR, the results of our GEMM experiments identify a  workload that is difficult to execute in a serverless environment. As shown in Figure~\ref{fig:gemm-eval}, 
{\proj} achieved worse performance than Dask. This is because GEMM natively requires a number of large data objects to be communicated between tasks before the computation can begin. Due to the limited bandwidth available to the intermediate data storage,
both {\proj} and numpywren experience long delays during this phase of the workload.

{\proj}'s performance greatly exceeded that of both numpywren configurations for all problem sizes.
For the largest problem size, {\proj} (single Redis shard) executed the workload 89.76\% faster than numpywren (single Redis shard). {\proj} (with Fargate) was 51.51\% faster than numpywren (with S3) for $15k \times 15k$ matrices because {\proj} reduced the amount of data read from and written to the intermediate KVS, and this reduction on I/Os directly translates to performance improvement. As shown in Figure~\ref{fig:gemm-io}, {\proj} read $49.39\%$ less data than numpywren for the smallest problem size and $45.24\%$ for the largest; {\proj} reduced the amount of data written by as much as $85\%$ for the largest problem size.

\begin{figure}
\vspace{-12pt}
\begin{minipage}{\textwidth}

\begin{minipage}[b]{0.26\textwidth}
\begin{center}
\includegraphics[width=1\textwidth]{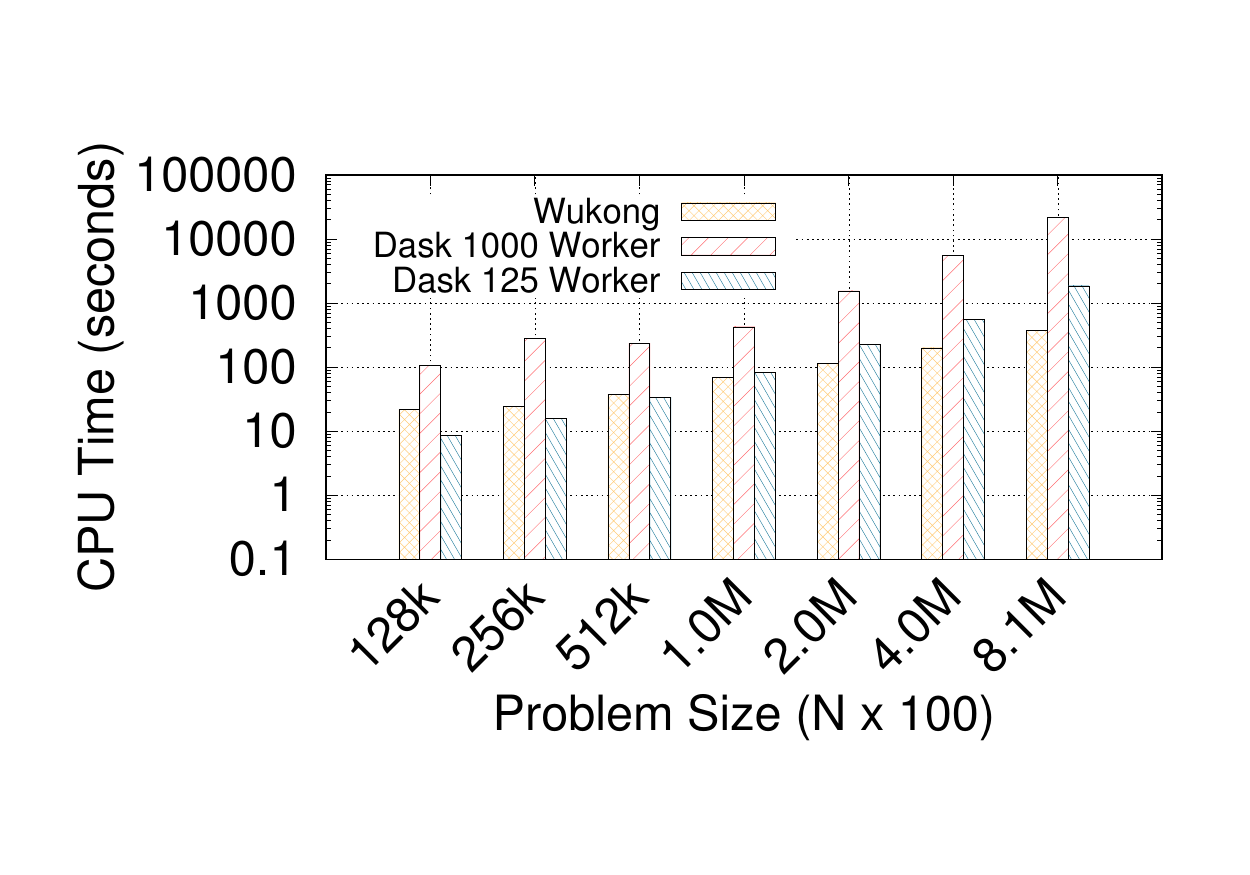}
\end{center}
\vspace{-20pt}
\caption{
\small{SVD1 CPU time.}}
\label{fig:svd1-core-seconds}
\end{minipage}
\hspace{-1em}%
\begin{minipage}[b]{0.26\textwidth}
\begin{center}
\includegraphics[width=1\textwidth]{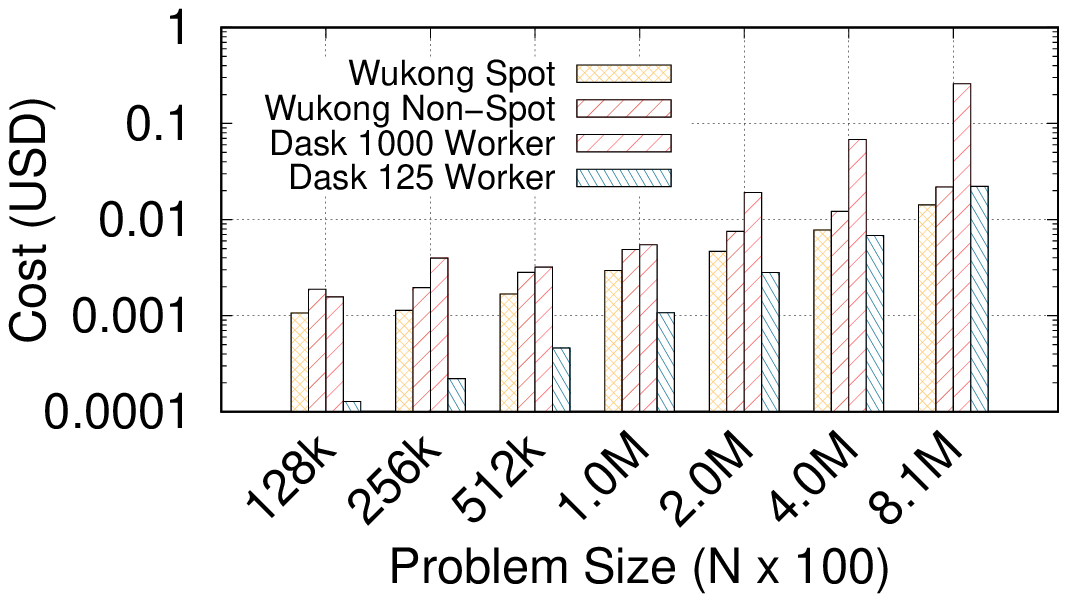}
\end{center}
\vspace{-20pt}
\caption{
\small{SVD1 cost.}}
\label{fig:svd1-cost}
\end{minipage}

\end{minipage}
\vspace{-15pt}
\end{figure}

\noindent\textbf{TSQR.} 
Figure~\ref{fig:tsqr-eval} shows that {\proj} outperformed both numpywren (with S3) and numpywren (single Redis shard) for all problem sizes. For the \(4.1M \times 128\) matrix, {\proj} (single Redis shard) was executing the workload \(68.17\times\) faster, or in 98.53\% less time, than numpywren (single Redis shard); and {\proj} (Fargate) is \(9.19\times\) faster, or in 89.11\% less time, than numpywren (S3). Numpywren (single Redis shard) failed to execute the next larger problem size, and the largest workload numpywren (S3) was able to execute was \(16.7M \times 128\). 
For this workload, {\proj} was \(13.36\times\) faster, executing the workload in $92.51\%$ less time. This again, is because of the significantly reduced reads and writes.  {\proj}'s ``become'' functionality allowed serial tasks along a single subgraph path to execute locally on the same Lambda Executor; whereas numpywren randomly assigned high priority tasks (which were ready to execute) to any stateless Lambda executor. As a result, numpywren wrote $16,027\times$ more data for the smallest problem size and $15,631\times$ more data for the largest problem size, which resulted in dramatic performance degradation (Figure~\ref{fig:tsqr-io}). 

\vspace{-4pt}
\subsection{CPU Time and Monetary Cost}
\label{subsec:cost}

Figure~\ref{fig:svd1-core-seconds} presents a comparison  between {\proj} (Multi-Redis) and both Dask clusters on their total CPU time (core seconds) for SVD1. Note that this comparison only considers cores actively used by Dask for each problem size.
{\proj} uses more core seconds than Dask-125 for the first three problem sizes, as Dask-125 finishes the job significantly faster than {\proj}. For $1.0M$ and above, {\proj} uses less core seconds than Dask-125. Dask-1000 incurs the longest CPU time as it is the slowest of the three. The disparity between the two frameworks increases as the problem size grows.

Figure~\ref{fig:svd1-cost} presents a comparison of the monetary cost to execute SVD1 for varying problem sizes.
This cost analysis only considers the Dask VMs actively used for each given workload size. 
At first, {\proj} is more expensive than Dask-125. As the problem size increases, the cost of executing the workload on {\proj} increases at a much slower rate than the cost of running the workload on Dask-125. By the largest problem size, the price of executing the workload on {\proj} is equal to that of executing the workload on Dask-125. Additionally, {\proj} achieves faster performance, lower cost, and more efficient CPU usage than Dask-1000 using non-spot pricing for all except the smallest problem size. These results demonstrate {\proj}'s pay-per-use property.

\begin{figure}[t]
\vspace{-10pt}
\begin{center}
\includegraphics[width=0.5\textwidth]{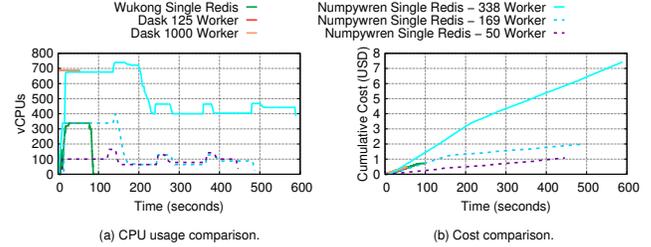}
\vspace{-25pt}
\caption{GEMM CPU usage and cost timeline.}
\vspace{-10pt}
\label{fig:gemm_cost}
\end{center}
\end{figure}

\begin{figure}[t]
\begin{center}
\includegraphics[width=0.5\textwidth]{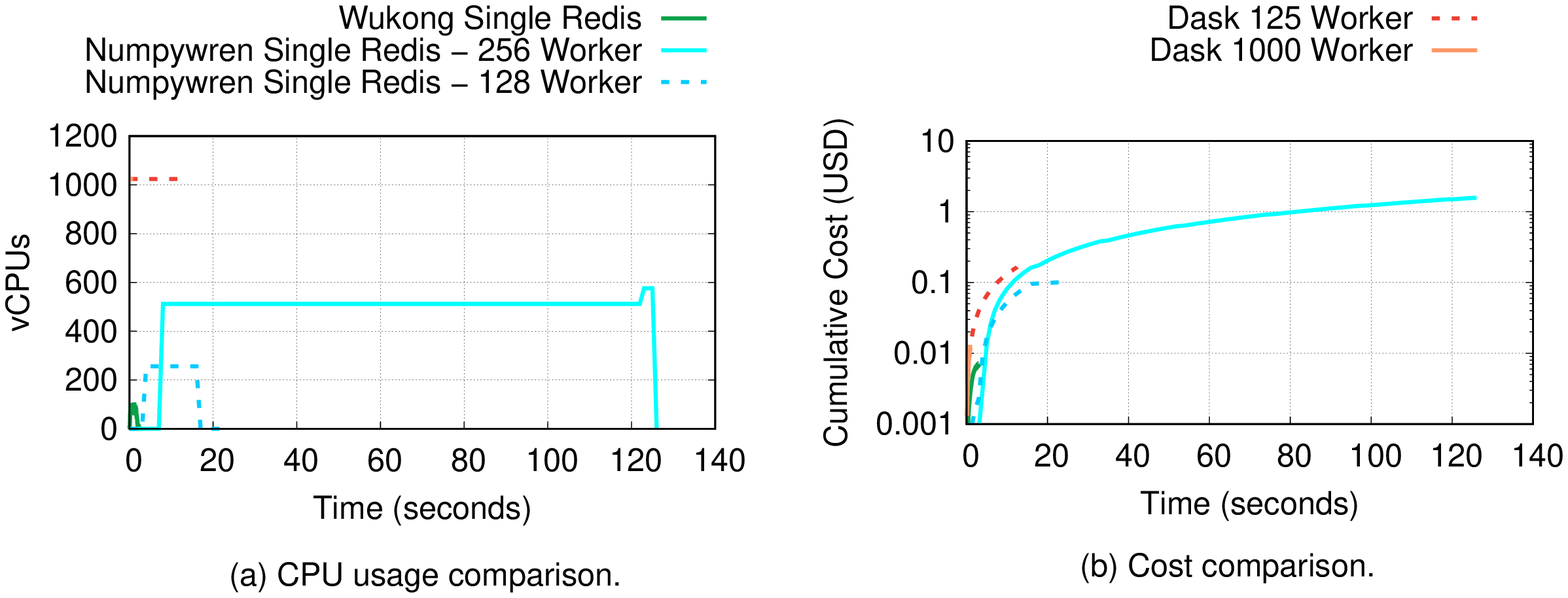}
\vspace{-25pt}
\caption{TSQR CPU usage and cost timeline.}
\vspace{-10pt}
\label{fig:tsqr_cost}
\end{center}
\end{figure}

\begin{figure*}[t]
\begin{center}
\vspace{-12pt}
\includegraphics[width=1\textwidth]{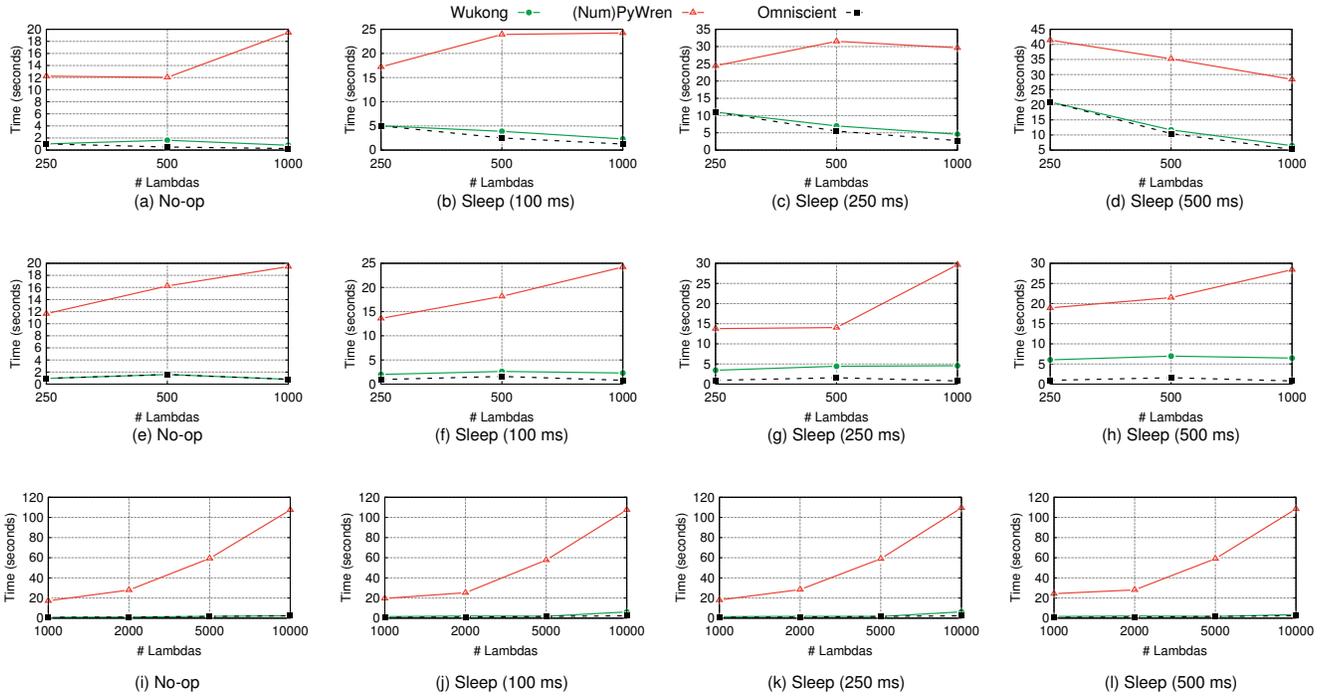}
\vspace{-60pt}
\caption{Figure~\ref{fig:scalability}(a)-(d) -- strong scaling: time (Y-axis) to execute $10,000$ tasks over $N$ Lambda executors (X-axis). 
Figure~\ref{fig:scalability}(e)-(h) -- weak scaling: time to execute $10$ tasks per Lambda. 
Figure~\ref{fig:scalability}(i)-(l) -- serverless scaling: time to execute $N$ task on $N$ Lambda.
For each row, plots are for (from left to right) tasks of 0, 100, 250, and 500~ms.}
\vspace{-10pt}
\label{fig:scalability}
\end{center}
\end{figure*}

Figure~\ref{fig:gemm_cost} shows a comparison between various configurations of {\proj} Single Redis, Dask, and numpywren Single Redis on the number of vCPUs and cumulative cost used during GEMM $25k \times 25k$. These configurations include both Dask configurations and three configurations of numpywren Single Redis. By design, numpywren allows users to specify the initial\footnote{
One can configure the starting number of executors, a maximum number of executors, and the policy used to scale the number of executors dynamically during execution. We opted to select the default scaling policy for all numpywren runs.
} number of Lambda executors (workers) for a job. We configured numpywren to use 50, 169, and 338 workers. Numpywren-169 was selected because the maximum concurrency achieved by {\proj} during this workload was 169 Lambdas. We tested a scaled-out version of numpywren that used twice as many starting workers (338) as well as a scaled-down version using 50 workers. Notably, numpywren-50 was the fastest configuration, followed by numpywren-169 and finally numpywren-338. This suggests that increasing the number of numpywren's parallelism significantly increases contention, possibly at the framework's central queue or scheduler, leading to performance degradation.

{\proj} was both cheaper and used less vCPUs during the execution of the workload than all three numpywren configurations. Specifically, {\proj} was $33.47\%$ cheaper and $77.57\%$ faster than the best-performing numpywren configuration. More importantly, {\proj} is autonomous and does not require users to explicitly tune the parallelism, which improves the usability.

Similarly, Figure~\ref{fig:tsqr_cost} shows a comparison for TSQR $4.0M$. The first configuration of numpywren used 128 workers while the second used 256 workers. These are based on the number of leaf tasks in {\proj}'s workload (512), though we found that using 512 numpywren workers resulted in worse performance. {\proj} used less CPU resources and was significantly cheaper than all other frameworks. It also out-performed all other frameworks except for Dask-125. Specifically, {\proj} completed the workload in $87.41\%$ less time and $92.96\%$ more cheaply than the best-performing numpywren configuration (i.e., $14.22\times$ cheaper and $7.94\times$ faster). Lastly, while {\proj} did not out-perform Dask-125, {\proj} was $95.67\%$ cheaper than Dask-1000 and $45.70\%$ cheaper than Dask-125 for this workload.
{\proj} did not reach more than 106 vCPUs during the workload's execution as many of the leaf tasks performed only a small amount of work before writing their data to the KVS and exiting. That is, by the time additional leaf tasks are scheduled, previously-invoked leaf tasks were already finishing their execution, forming a scheduling pipeline.

\vspace{-15pt}
\subsection{Elastic Scaling}
\label{subsec:scaling}

We next evaluate {\proj}'s scalability on traditional strong / weaking scaling and serverless scaling metrics, and compare it against (Num)-PyWren. The maximum concurrency we got from Amazon was $5,000$ concurrent Lambdas. In this experiment, for both (Num)-PyWren and {\proj}, all the executors in the Lambda pool got warmed up before accepting task requests, eliminating the cold start concerns. The Lambda pool scaled from zero.

\noindent\textbf{Strong Scaling.} 
For strong scaling, we varied the number of Lambdas to be launched to execute $10,000$ tasks. In order to simulate workloads with various computation loads, we added a delay of 100~ms, 250~ms, and finally 500~ms to each task. Each test was repeated three times. Figure~\ref{fig:scalability}(a)--(d) show that, in all cases, {\proj} exhibited near-ideal strong scaling behavior, scaling significantly better than (Num)-PyWren in all cases, thereby demonstrating the effectiveness of our decentralized scheduling mechanism. It was not until the 500~ms delay case that (Num)-PyWren exhibited a downward trend in execution time as the number of Lambda executors scaled. This is because: 1) 500-ms tasks tend to run longer, and 2) with more Lambda executors and a fixed total amount of tasks, each Lambda gets assigned less number of tasks.

\noindent\textbf{Weak Scaling.}
For weak scaling, we executed ten tasks per worker and varied the number of Lambda executors from $250$ to $1,000$. 
As shown in Figure~\ref{fig:scalability}(e)--(h), {\proj} exhibited near-ideal weak scaling behavior for all sleep delays and worker configurations. Notably, thanks to the decentralized scheduling, {\proj} was able to scale to $1,000$ Lambda executors much more effectively than (Num)PyWren, which experienced increasingly large delays as the number of Lambdas increased. 

\noindent\textbf{Serverless Scaling.}
Finally, we test serverless scaling -- executing $N$ tasks on $N$ Lambda executors, with each Lambda effectively executing one single task. Figure~\ref{fig:scalability}(i)--(l) plots the results. We observe that (Num)-PyWren took an increasing amount of time to finish executing $N$ tasks on $N$ Lambdas, and executing $10,000$ tasks took almost two minutes. In contrast, {\proj} rapidly scales out to $10,000$ tasks in few seconds, almost approaching the behavior of an omniscient serverless execution framework. This again demonstrates the efficacy of {\proj}'s decentralized scheduling.

\begin{figure}[t]
\begin{center}
\includegraphics[width=0.48\textwidth]{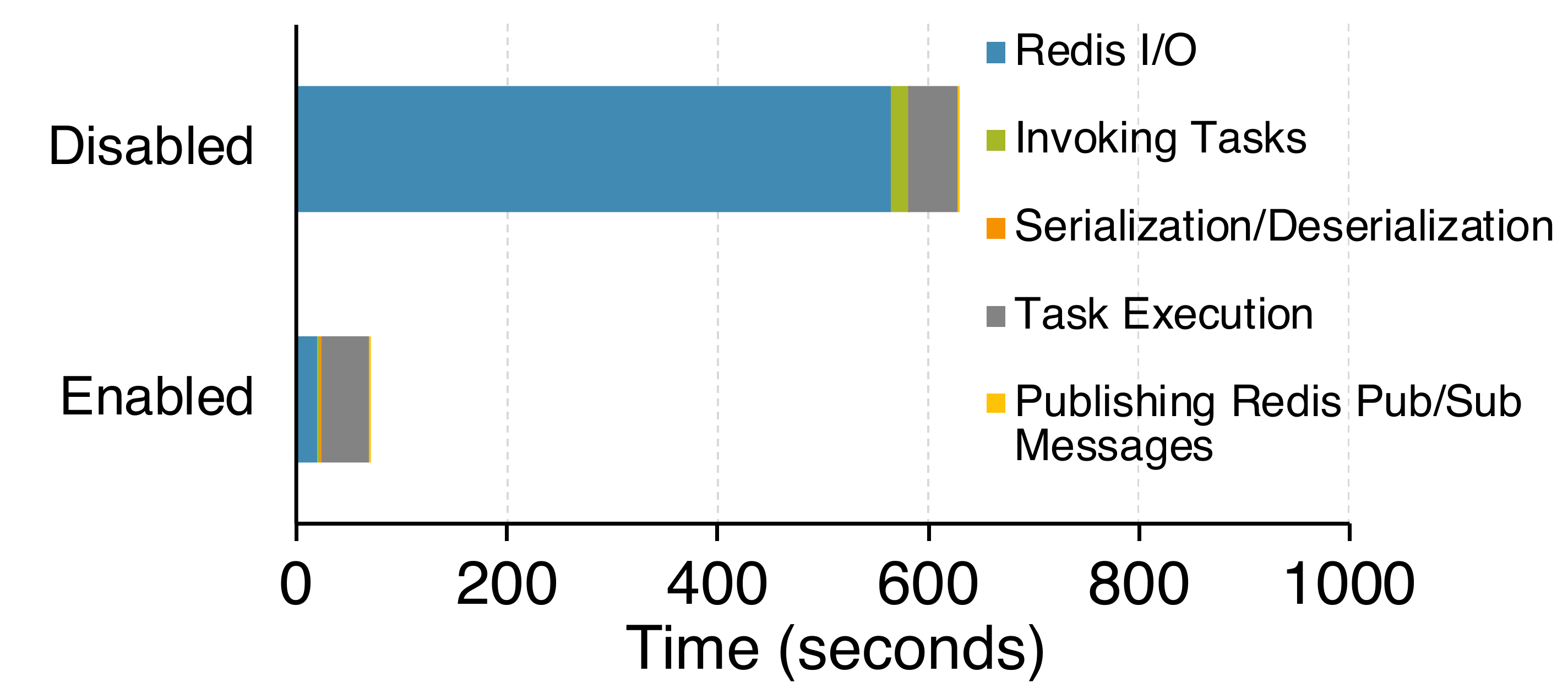}
\vspace{-10pt}
\caption{
SVD2 50k x 50k aggregated execution time breakdown with and without task clustering and delay I/O.
}
\vspace{-10pt}
\label{fig:svd-cluster-vs-no-cluster}
\end{center}
\end{figure}

\vspace{-6pt}
\subsection{Factor Analysis}
\label{subsec:factor}

\noindent\textbf{Task Clustering and Delay I/O.} Finally, we look at the impact of task clustering and delaying I/O on {\proj}'s performance.
Clustering and delay I/O prevent tasks with large intermediate data from writing their data to Redis. Instead, these tasks attempt to execute downstream tasks locally. Any downstream tasks that cannot immediately be executed because their dependencies are not satisfied are put into a queue. The dependencies of the queued tasks are routinely rechecked until the dependencies are satisfied or a maximum delay time is reached. By delaying I/O, more tasks can be clustered and expensive network I/Os data can be avoided, decreasing the workload runtime dramatically. 

\begin{figure}[t]
\begin{center}
\vspace{-25pt}
\includegraphics[width=0.45\textwidth]{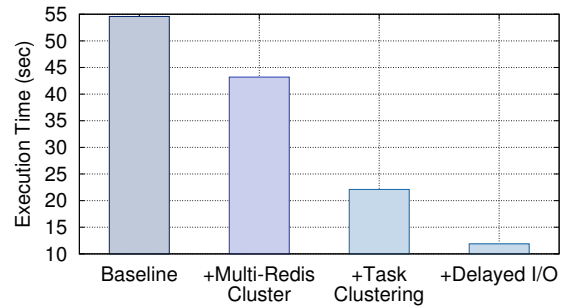}
\vspace{-25pt}
\caption{
Contributions of optimizations to {\proj}'s performance of SVD2.
}
\vspace{-20pt}
\label{fig:abc-optimizations}
\end{center}
\end{figure}

Figure~\ref{fig:svd-cluster-vs-no-cluster} displays two aggregations of time for the activities performed for SVD2. 
In both cases, ``publishing messages'', ``task execution'', and ``serialization/deserialization'' each took roughly the same amount of time (in aggregate); however, the difference between the times for task invocation and especially Redis I/O is significant. With the two optimizations disabled, task invocations and Redis I/O made up an aggregate $14.80$ and $565.21$ seconds, respectively. When the optimizations were enabled, task invocation took an aggregate $2.05$ seconds while Redis I/O took just $20.36$ seconds. There is \(7.21\times\) more aggregate time spent invoking tasks and \(27.76\times\) more aggregate network I/O performed with clustering disabled. 

We analyze {\proj}'s performance by breaking down the performance gap between a baseline and {\proj} with all optimizations enabled (Figure~\ref{fig:abc-optimizations}).
The use of the Fargate multi-Redis storage cluster results in a $20.85\%$ performance improvement over using AWS ElastiCache for intermediate data storage. When using Fargate, the I/O performed during the workload is spread across a large number of Redis instances, resulting in reduced network contention and consequently reduced I/O latency. (Using a large number of ElastiCache instances is cost prohibitive.) When clustering (without delayed I/O) is enabled, performance improves by another $48.82\%$ as a significant amount of large object I/Os is eliminated. Finally, enabling delayed I/O results in a $46.21\%$ improvement relative to the use of clustering and Fargate alone. Overall, {\proj} is $4.6\times$ faster when all optimizations are used, demonstrating the effectiveness of these techniques.
\vspace{-10pt}
\section{Related Work}
\label{sec:design}
 
Researchers have identified new stateful parallel  applications for serverless computing. 
These efforts have lead to serverless parallel computing frameworks~\cite{pywren_socc17, numpywren, excamera_nsdi17, sprocket_socc18, crucial_middleware19, cirrus_socc19, step_functions, HyperFlow, locus_nsdi19, gg_atc19, dl_serving_ic2e18}, which have been built using methods 2 and 3 describes in \cref{sec:intro}.
However, none of them explicitly addresses the data locality issue of stateful serverless parallel applications. {\proj} goes beyond existing work with a novel \emph{locality-aware} decentralized scheduling approach for complex DAG jobs. 

funcX~\cite{funcx_hpdc20} is an open FaaS platform that enables high performance ``serverless supercomputing'' over existing HPC infrastructures. {\proj} is orthagonal to funcX and should be portable to existing commercial and open-source FaaS platforms ~\cite{fission, openfaas, openwhisk, google_func, azure_func}. Porting {\proj} to other serverless platforms is part of our future work.

SAND~\cite{akkus_sand_2018} is a serverless platform that increases data locality by running some or all of the functions/tasks for a given workload within the same container on the same server, but as separate processes. 
{\proj} is a serverless application framework that increases data locality by executing multiple dependent tasks in the same {\proj} Executor function, and hence in the same process, container, and server. Thus, SAND improves data locality in the serverless platform layer, while {\proj} improves locality in the application framework layer running atop the serverless platform.

\textsc{InfiniCache}~\cite{infinicache_fast20} is a distributed memory cache that exploits the memory of serverless functions for object caching. {\proj} complements \textsc{InfiniCache} in that {\proj} can use \textsc{InfiniCache} to cache intermediate data. 

General-purpose serverless orchestration frameworks include AWS Step Functions~\cite{step_functions}, Azure Durable Functions~\cite{azure_durable_func}, Fission Workflows~\cite{fission_workflows}, and HyperFlow~\cite{malawski_serverless_hyperflow_2017, HyperFlow}. But these frameworks are not well-suited for supporting large, complex jobs, because they require manual workflow configurations (e.g., JSON)~\cite{lopez_comparison_2018}.

A large body of research has explored distributed scheduling~\cite{ray_osdi18, sparrow_sosp13, omega_eurosys13, dague_pc12, parsl_hpdc19}.
However, these solutions all target serverful scheduling with serverful deployment specific optimization objectives.
{\proj} targets serverless computing and takes advantage of the massive parallelism of serverless computing for high efficiency.

\vspace{-4pt}
\section{Discussion and Lessons}
\label{sec:discussion}

{\proj} is not intended to replace established, serverful processing frameworks such as Spark~\cite{spark_nsdi12} and TensorFlow~\cite{tensorflow_osdi16}. Rather, because {\proj} is less powerful but easier to use than these serverful frameworks, {\proj}
targets users who lack a strong CS background, but who work on lighter-weight and smaller computing-related problems, in areas such as data analytics and machine learning, particularly those whose solutions can be implemented using numerical Python libraries.

Our initial attempt to port Dask to a serverless platform was unsuccessful due to the poor performance of the resulting framework~\cite{wukong_pdsw19}. There were several major bottlenecks in this original design. For one, the use of a centralized scheduler to assign tasks to Lambda executors was too slow for all but the smallest workloads. Specifically, the centralized scheduler was unable to process thousands of concurrent network connections with Lambda executors. Additionally, the centralized scheduler struggled to rapidly invoke Lambda functions when scaling out for large workloads. Finally, Lambda executors spent an overwhelming majority of their time reading and writing intermediate data. \cite{wukong_pdsw19} presents a preliminary study about these performance bottlenecks.

To address these performance issues, we developed our decentralized scheduling technique. While this technique significantly improved performance, there were still bottlenecks due to a lack of data locality. Specifically, reading and writing very large intermediate objects dominated end-to-end execution time, particularly for workloads such as SVD. To address these bottlenecks, we developed task clustering and delayed I/O. Though these techniques have been used in serverful contexts~\cite{kwok_acmcs, dague_pc12}, they had never been utilized in a serverless context. With the addition of these two techniques, large object reads and writes were eliminated, which greatly improved performance and resource utilization.

\vspace{-4pt}
\section{Conclusion}
\label{sec:conclusion}

We have presented {\proj}, a new serverless parallel computing framework that uses locality-enhanced, decentralized scheduling (atop AWS Lambda), task clustering, and delayed I/O
to achieve high performance, near-ideal scalability, and data locality while being cost-effective.
Our evaluation demonstrates the effectiveness of decentralized scheduling, task clustering, and delayed I/O in reducing both the execution time and cost of executing workloads. Further, we have shown that {\proj} exhibits near-ideal scaling behavior, reduces the execution time by as much as $98.53\%$ compared to numpywren, and reduces network I/O by multiple orders of magnitude. 
Finally, {\proj} can reduce costs by upwards of $92.96\%$ and $95.67\%$ compared to numpywren and Dask, respectively.

{\proj}'s source code is available at:
\begin{center}
\textbf{\url{https://mason-leap-lab.github.io/Wukong}}.
\end{center}

\label{startofrefs}
\clearpage
\newpage

\section*{Acknowledgments}

We are grateful to the anonymous reviewers for their valuable comments and suggestions that improved the paper. This work is sponsored in part by NSF grants CCF-1919075, CCF-1919113, OAC-2007976, George Mason University, an AWS Cloud Research Grant, and Google Cloud Platform Research Credits.

\bibliographystyle{ACM-Reference-Format}
\bibliography{socc}

\newpage

\end{document}
\endinput